%% file: ms.tex
\documentclass[iop]{emulateapj}

\usepackage{amsmath}    
\usepackage{natbib}
\usepackage{color}
\usepackage{graphicx}
\usepackage{epstopdf}
\received{}
\accepted{}

\slugcomment{to appear in ApJ}
\shorttitle{}
\shortauthors{Wang et al. - HDC}

\begin{document}
\title{Observing Exoplanets with High Dispersion Coronagraphy.\\
I. The scientific potential of current and next-generation large ground and space telescopes}

\author{
Ji Wang\altaffilmark{1},
Dimitri Mawet\altaffilmark{1},
Garreth Ruane\altaffilmark{1},
Renyu Hu\altaffilmark{2, 3}, and
Bj{\"o}rn Benneke\altaffilmark{3}
} 
\email{ji.wang@caltech.edu}
\altaffiltext{1}{Department of Astronomy, California Institute of Technology, MC 249-17, 1200 E. California Blv, Pasadena, CA 91106 USA}
\altaffiltext{2}{Jet Propulsion Laboratory, California Institute of Technology, Pasadena, CA 91109, USA}
\altaffiltext{3}{Division of Geological and Planetary Sciences, California Institute of Technology, Pasadena, CA 91125, USA}

\begin{abstract}

Direct imaging of exoplanets presents a formidable technical challenge owing to the small angular separation and high contrast between exoplanets and their host stars. High Dispersion Coronagraphy (HDC) is a pathway to achieve unprecedented sensitivity to Earth-like planets in the habitable zone. Here, we present a framework to simulate HDC observations and data analyses. The goal of these simulations is to perform a detailed analysis of the trade-off between raw star light suppression and spectral resolution for various instrument configurations, target types, and science cases. We predict the performance of an HDC instrument at Keck observatory for characterizing directly imaged gas-giant planets in near infrared bands. We also simulate HDC observations of an Earth-like planet using next-generation ground-based (TMT) and spaced-base telescopes (HabEx and LUVOIR). We conclude that ground-based ELTs are more suitable for HDC observations of an Earth-like planet than future space-based missions owing to the considerable difference in collecting area. For ground-based telescopes, HDC observations can detect an Earth-like planet in the habitable zone around an M dwarf star at 10$^{-4}$ starlight suppression level. Compared to the 10$^{-7}$ planet/star contrast, HDC relaxes the starlight suppression requirement by a factor of 10$^3$. For space-based telescopes, detector noise will be a major limitation at spectral resolutions higher than 10$^4$. Considering detector noise and speckle chromatic noise, R=400 (1600) is the optimal spectral resolutions for HabEx(LUVOIR). The corresponding starlight suppression requirement to detect a planet with planet/star contrast=$6.1\times10^{-11}$ is relaxed by a factor of 10 (100) for HabEx (LUVOIR).

\end{abstract}


\section{Introduction}
\label{sec:intro}

Out of the thousands of exoplanets detected to date, the few that have been directly imaged are excellent targets for studying orbital configurations~\citep{Pueyo2015, Zurlo2015, Blanchaer2015, Maire2015} and atmospheric chemical compositions~\citep{Konopacky2013, Oppenheimer2013, Bonnefoy2015, Rajan2015}. However, direct imaging and characterization faces several technical challenges owing to the small angular separation and high contrast between exoplanets and their host stars. High-contrast imaging (HCI) systems mitigate these effects by suppressing diffracted star light, that may otherwise overwhelm the planet signal, with an extreme adaptive optics system and a coronagraph. Current state-of-the-art high contrast imaging instruments, such as the Gemini Planet Imager at the Gemini South telescope~\citep{Macintosh2014} and SPHERE at the Very Large Telescope~\citep{Beuzit2008}, are able to achieve better than $10^{-4}$ star light suppression level at a few tenths of an arcsec, which allows for the detection of gas giant planets and brown dwarfs orbiting nearby young stars~\citep[e.g., ][]{Macintosh2015, Wagner2016}. 

Star light suppression can be further improved by coupling a high-resolution spectrograph (HRS) with a coronagraphic system~\citep{Sparks2002, Riaud2007, Kawahara2014, Snellen2015, Lovis2016}. In this High Dispersion Coronagraphy (HDC) scheme, the coronagraphic component serves as a spatial filter to separate the light from the star and the planet. The HRS serves as spectral filter taking advantage of differences in spectral features between the stellar spectrum and the planetary spectrum, e.g., different absorption lines and radial velocities (RV). 

Using HRS as a way of spectral filtering has been successfully demonstrated by a number of teams. For example, high-resolution transmission spectroscopy has been used to detect molecular gas in the atmospheres of transiting planets~\citep{Snellen2010, Birkby2013, deKok2013}. At a high spectral resolution, resolved molecular lines can be used to study day- to night-side wind velocity~\citep{Snellen2010} and validate 3D exoplanet atmosphere models~\citep{Kempton2014}. For planets detected via RV, the spectral lines due to the planet can be separated from stellar lines with their drastically different RVs ($\gtrsim$ 50 km s${^{-1}}$). Thus, the RV of the planet itself may be measured to break the degeneracy between the true planet mass and orbital inclination~\citep{Brogi2012, Brogi2013, Brogi2014, Lockwood2014}. Moreover, HRS permits detailed study of spectral lines arising from a planet's atmosphere. This approach led to the first measurement of a planet's rotational velocity~\citep{Snellen2014}. With time-series HRS, surface features such as cloud or spot coverage may be inferred via Doppler imaging, which has been demonstrated on the closest brown dwarf system, Luhman 16 AB~\citep{Crossfield2014}. 

As showcased by the examples above, HRS may be used to detect planets that are $\sim10^{-4}$ times as bright as their host stars. When coupled with a state-of-the-art HCI system capable of reaching star light suppression levels of $\sim10^{-4}$, an HDC instrument is sensitive to much fainter planets. Meanwhile, relatively bright planets may be observed at a higher signal-to-noise ratio (SNR) allowing for the physical and chemical processes taking place in their atmospheres to be studied in greater detail. Here, we develop a framework to simulate the performance of an HDC instrument. Although similar calculations have been performed as part of previous studies~\citep{Sparks2002, Riaud2007, Kawahara2014, Snellen2015, Lovis2016}, a thorough end-to-end simulation that explores the SNR trade space between spectral resolution and starlight suppression for ground-based and space-based observations is lacking. In this paper, we simulate a variety of HDC instruments that are either under development or in the conceptual design phase and quantify their potential for detecting new planets as well as particular molecular species in the atmosphere of known planets (e.g. Proxima Cen b, 51 Eri b, HR 8799 e) and hypothetical Earth-like planets around stars of different spectral types. 

The paper is organized as follows. We outline the procedure used to simulate the performance of an HDC instrument for detecting and characterizing exoplanets in \S \ref{sec:simulation}. The planned Keck HDC instrument is briefly described in \S \ref{sec:instrument}. We study the prospects of using the Keck HDC instrument to observe previously imaged exoplanets in \S \ref{sec:ScienceCases}. HDC observations of potential Earth-like planets (e.g. Proxima Cen b) in the habitable zone of M dwarfs are investigated in \S \ref{sec:Earths} for current and next-generation extremely large telescopes. Observing Earth-like planets around solar-type stars with future space telescopes is considered in \S \ref{sec:sun_earth}. A summary and discussion are provided in \S \ref{sec:discussion}.

\section{HDC fundamental trade-off analysis}
\label{sec:simulation}
\subsection{Simulating the Observations}
\label{sec:simulation_observation}

\begin{figure*}
\epsscale{1.0}
\plotone{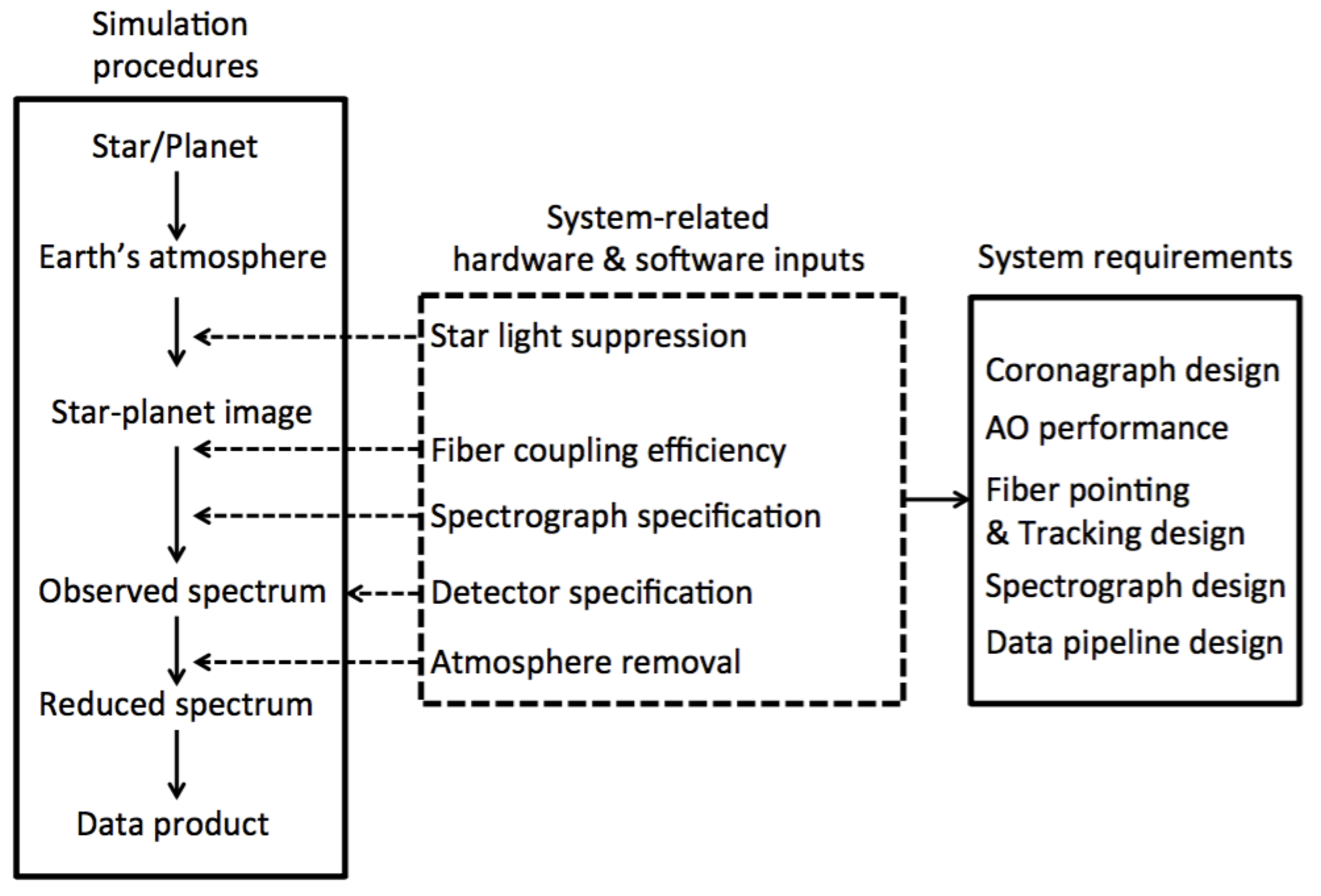}
\caption{Flow chart of simulation of an HDC instrument. Photons from star and planet go through an HCI instrument. Residual star light and planet light is picked up by a single-mode fiber, which feeds the light into an HRS instrument. Spectra are simulated with detector noise and then reduced into data product, e.g., CCF (see \S \ref{sec:data_reduction}). Atmospheric effect is optional depending on ground-based or space-based observation. The simulation pipeline provides a way of setting system requirements for an HDC instrument and understanding the fundamental limit of the HDC technique. 
\label{fig:flow_chart}}
\end{figure*}

In this section, we describe our workflow to simulate the end-to-end performance of an HDC system, from the intrinsic spectrum of a planet and star to the measured spectrum and the subsequent post-processing. The goal of these simulations is to perform a detailed analysis of the trade-off between raw star light suppression and spectral resolution for various instrument configurations, target types, and science cases.
Fig.~\ref{fig:flow_chart} shows a flow chart to illustrate the procedure and the system-related inputs to the simulation. The resulting data products, e.g., cross correlation fuction (CCF) and their quality (e.g. SNR) will inform observation strategies and system requirements, including the coronagraph design and the performance of the adaptive optics (AO) system. 

\subsubsection{Generating Spectra of Stars and Planets}
\label{sec:simulation_spec}

Gas-giant planet spectra consisting of all molecular species are derived from the published BT-Settl grids\footnotetext{https://phoenix.ens-lyon.fr/Grids/BT-Settl/CIFIST2011\_2015/FITS/}~\citep{Baraffe2015}. The grids cover effective temperatures ($T_{\rm{eff}}$) from 1200 K to 7000 K. For $T_{\rm{eff}}$ outside of this range, we use the BT-Settl grids with \citet{Caffau2010} solar abundances\footnote{https://phoenix.ens-lyon.fr/Grids/BT-Settl/CIFIST2011/SPECTRA/} (400 K $<T_{\rm{eff}}<$ 8000 K). The stellar spectra used in our simulations are also derived from these grids, which cover the $T_{\rm{eff}}$ and log(g) range of host stars considered here. If necessary, the planet and star fluxes are scaled to match the observed absolute flux.

High-resolution spectra discerning the individual contributions of the molecular absorbers $\mathrm{H_{2}O}$, CO, and $\mathrm{CH_{4}}$ are simulated using the SCARLET model ~\citep{Benneke2015, Benneke2013}. In this work, SCARLET first iteratively computes the line-by-line radiative transfer and atmospheric chemistry to converge to a self-consistent vertical temperature structure and molecular composition. To isolate the contribution from individual molecules, we then artificially remove all opacities in the atmosphere except the opacity of the respective molecular absorber and collision-induced absorption in the simulation of the planets' thermal emission spectra. In this way, we compute emission spectra for each of the molecular absorbers individually. 
The SCARLET model considers the molecular opacities of $\mathrm{H_{2}O}$, $\mathrm{CH_{4}}$, $\mathrm{NH_{3}}$, HCN, $\mathrm{CO}$, and $\mathrm{CO_{2}}$ and $\mathrm{TiO}$ from the high-temperature ExoMol database \citep{Tennyson2012}, and $\mathrm{O_{2}}$, $\mathrm{O_{3}}$, $\mathrm{OH}$, $\mathrm{C_{2}H_{2}}$, $\mathrm{C_{2}H_{4}}$, $\mathrm{C_{2}H_{6}}$, $\mathrm{H_{2}O_{2}}$, and $\mathrm{HO_{2}}$ from the HITRAN database \citep{Rothman2009}. Absorption by the alkali metals (Li, Na, K, Rb, and Cs) is modeled based on the line strengths provided in the VALD database \citep{Piskunov1995} and $\mathrm{H_{2}}$-broadening prescription provided in \citet{Burrows2003}. Collision-induced broadening from $\mathrm{H_{2}}/\mathrm{H_{2}}$ and $\mathrm{H_{2}/He}$ collisions is computed following \citet{Borysow2002}.

The spectra of Earth-like exoplanets, on the other hand, are generated by an atmospheric chemistry and radiative transfer model~\citep{Hu2012a, Hu2012b, Hu2013, Hu2014}. We first calculate the molecular abundance as a function of altitude, controlled by photochemical and disequilibrium chemistry processes. The details of the model are described in~\citet{Hu2012a} and the molecular abundances have been validated against measurements on Earth. 

We include the effects of clouds in the resulting spectra by averaging two scenarios: a cloud-free scenario where we assume a clear atmosphere and a high-cloud scenario where we assume a reflective H$_2$O cloud at 9-13 km. This procedure produces a continuum albedo of $\sim$0.3 and provides a realistic estimate of the strength of spectral features, similar to~\citet{DesMarais2002}. Eighth-order Gaussian integration is used to calculate the contribution of the whole planetary disk for both the reflected light and thermal emission. We include the opacities of CO$_2$, O$_2$, H$_2$O, CH$_4$, O$_3$, and N$_2$O and calculate the planetary flux at a spectral resolution of $R=\lambda/\Delta\lambda=500,000$, high enough to resolve individual spectral lines of the aforementioned species over $\lambda=0.5$-5 $\mu$m. The resulting spectra are then expressed as albedo and scaled with the planet's size within the reasonable range for terrestrial planets.

\subsubsection{Spectrum of Earth's Atmosphere}

Telluric and sky emission lines are included in the simulation to account for additional photon loss, near infrared background noise, and potential confusion between molecules that appear in both the planet's and Earth's atmosphere, e.g., H$_2$O and O$_2$. We use the Mauna Kea sky transmission\footnote{http://www.gemini.edu/sciops/telescopes-and-sites/observing-condition-constraints/ir-transmission-spectra} and emission spectra\footnote{http://www.gemini.edu/sciops/telescopes-and-sites/observing-condition-constraints/ir-background-spectra}, available from the Gemini observatory website~\citep{Lord1992}, with wavelength coverage of 0.9-5.6 $\mu$m. A water column density of 1.6 mm and airmass of 1.5 is assumed. Since we also consider telluric absorption at shorter wavelengths, we also use telluric absorption data from the National Solar Observatory for wavelengths shorter than 0.9~$\mu$m\footnote{diglib.nso.edu/ftp.html}. 

\subsubsection{Simulation Procedure}


An HDC instrument contains two major components, a coronagraph and a high-resolution spectrograph, which are linked by a set of single-mode fibers: a planet fiber, a star fiber, and/or a sky fiber. One end of these fibers is located at the image plane after the coronagraph and the other end of the fibers is at the entrance slit of the spectrograph. The star fiber and sky fiber provide calibration spectra in data reduction described in \S \ref{sec:data_reduction}.  Following Fig. \ref{fig:flow_chart}, light from the star and planet go through a coronagraph and form an image. The fiber at the planet location leads the planet light, as well as residual star light, into the spectrograph. The detector records the planet spectrum along with contaminating stellar spectrum. We note that atmospheric effect, i.e., absorption and emission, is only considered for ground-based observations.

\begin{equation}
\label{eq:f_observed}
\rm{f_{\rm{detector}}} = (f_{\rm{planet}}+f_{\rm{star}}\times C)\times f_{\rm{transmission}} + f_{\rm{sky}}.
\end{equation}

We simulate the signal recorded on a detector as described in Equation \ref{eq:f_observed}. Flux from a star and a planet is in the unit of W$\cdot\rm{m}^{-2}\cdot\mu\rm{m}^{-1}$ at a reference height ($d_{\rm{ref}}$). We calculate the incident star and planet photon flux on the detector, $f_{\rm{star}}$ and $f_{\rm{planet}}$, with the following equation: $f=F\times(d_{\rm{ref}}/d)^{2}\times{\rm{A}}\times\Delta\lambda\times\eta\times{\rm{t}}_{\rm{exp}}\ /\ h\nu$, where $F$ is the flux at the reference height ($d_{\rm{ref}}$), $d$ is distance between the star-planet system and an observer, A is telescope receiving area, $\Delta\lambda$ is wavelength coverage per wavelength bin, $\eta$ is telescope and instrument end-to-end throughput, t$_{\rm{exp}}$ is exposure time, $h$ is Planck constant and $\nu$ is the frequency of a photon. 

At the image plane after a coronagraph, stellar flux is suppressed by a factor $C$, a parameter we denote as star light suppression factor, i.e., the fraction of the total starlight that couples into the fiber.

Both stellar and planetary spectra are rotationally broadened. The effect of rotation broadening is calculated by summing spectra from surface grids evenly spaced in longitudinal and latitudinal direction. The rotationally-broadened spectra are then multiplied by the Earth's atmosphere transmission spectrum ($f_{\rm{transmission}}$) for ground-based observation. The Earth's atmosphere transmission spectrum is unitless and normalized to unity, with zero meaning entirely opaque and one meaning entirely transmissive. For space-based observations, $f_{\rm{transmission}}$ is set to unity.

The spectra are then broadened by instrumental line spread function (LSF).  The instrumental broadening is approximated by convolving a spectrum with a normalized Gaussian core with a full width at half maximum (FWHM) of one spectral resolution element, which is $\lambda_0$/R, where $\lambda_0$ is central wavelength and R is the spectral resolution of a spectrograph. 

The broaden spectra are then added by sky emission spectrum ($F_{\rm{emission}}$), which is also broadened at a given spectral resolution. The sky emission is in the unit of photons$\cdot{\rm{s}}^{-1}\cdot{\rm{arcsec}}^{-2}\cdot{\rm{m}}^{-2}\cdot\mu{\rm{m}}^{-1}$. We calculate the incident photon flux from sky emission on detector using the following equation: $f_{\rm{emission}} = F_{\rm{emission}}\times{\rm{t}}_{\rm{exp}}\times\theta^{2}\times{\rm{A}}\times\Delta\lambda$, where $\theta^{2}$ is the projected area of sky to the input fiber fundamental mode size which we assume to be $(1.0\ \lambda_0/\rm{D})^2$, where $\lambda_0$ is central wavelength and D is telescope aperture diameter. For space-based observation, $f_{\rm{emission}}$ is set to zero. The simulated spectra are then resampled at the pixel sampling rate per resolution element.  

In addition to the spectrum described by Equation \ref{eq:f_observed}, we simulate more spectra for subsequent data reduction. For ground-based observation, we simulate sky emission and stellar spectra, assuming there are two dedicated fibers for sky and star. The stellar spectrum can be used to remove atmospheric transmission and/or contaminated stellar lines in the planet spectrum. For example, in the case of ground-based observations of the HR 8799 system, the host star itself is a fast rotating early-type star and thus can be used as a telluric standard to remove atmosphere transmission. In the case of ground-based observations of Proxima Cen b, the observed spectrum is a reflection spectrum containing both the star and planet absorption lines and is contaminated by the Earth's atmosphere lines, so it is necessary to have a separate simultaneous observation of the host star to remove atmospheric transmission and/or contaminated stellar lines in the planet spectrum. For space-based observations, we simulate only the stellar spectrum since the background is negligible. 

\subsubsection{Noise Sources}

We include realistic estimates of photon noise, detector readout noise, and dark current based on the performance of a Teledyne HgCdTe H2RG infrared detector and a e2v optical charge-coupled device (CCD). Readout noise and dark current for the H2RG detector are 2.0 e$^{-}$ (Fowler-32 readout, personal correspondence with Roger Smith) and 0.002 e$^{-}/s$~\citep{Blank2012}, respectively. An e2v optical CCD\footnote{http://www.e2v.com/resources/account/download-datasheet/1364} has a readout noise of 2.0 e$^{-}$ and a dark current of 0.02 e$^{-}/\rm{hour}$.

\begin{equation}
\label{eq:f_noise}
\delta = \sqrt{\rm{f} + \rm{n}_{\rm{exp}} \times RN^{2} + \rm{dark}\times t_{\rm{exp}} },
\end{equation}

The total noise is calculated by Equation \ref{eq:f_noise}, where $\delta$ is the combined noise, f is the photon noise followed by terms for readout noise (RN) and dark current (dark), $\rm{n}_{\rm{exp}}$ is the number of readout within a total observation time $t_{\rm{exp}}$. The number of readouts $\rm{n}_{\rm{exp}}$ is determined by the linear range or detector persistence limit, i.e. the signal level where a new frame needs to be taken in order to avoid non-linearity or persistence. Exposure time per frame or the number of readouts is usually set by the raw level of star light suppression (instrumental contrast) or sky background emission. We make a conservative assumption that the persistence limit is at 12000 electrons. During operations such as sky emission removal, telluric/stellar line removal, noises are added in quadrature.

\subsection{Spectral Analyses}
\label{sec:data_reduction}

\begin{figure*}
\epsscale{1.2}
\plotone{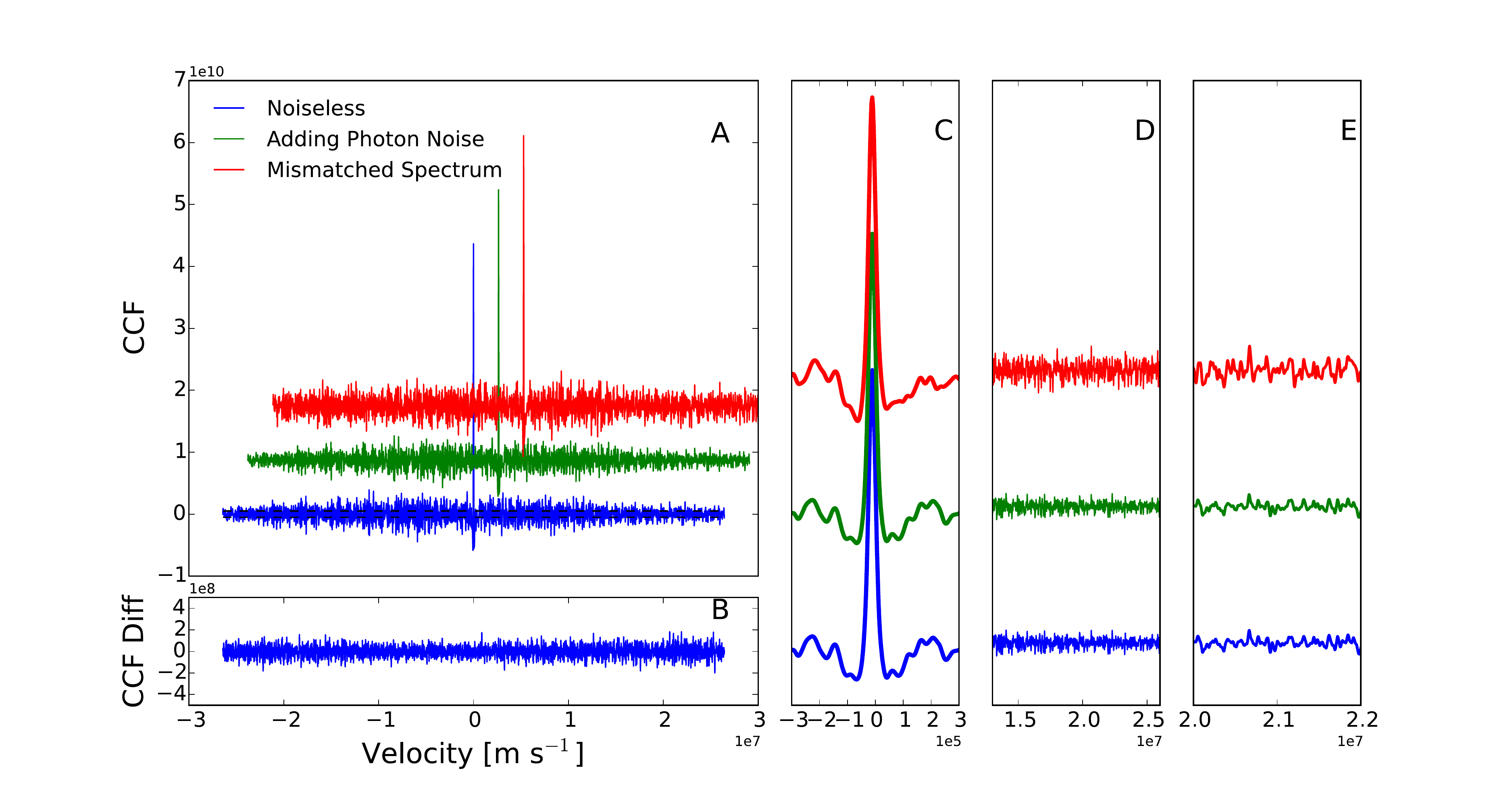}
\caption{A: examples of cross correlation function (see \S \ref{sec:HR8799e} for definition of different cases). CCFs are vertically and horizontally offset for visual clarity. Peaks of CCFs are scaled to the same height to emphasize different fluctuation level outside CCF peaks. Dashed lines indicate lower and upper boundaries for y-axis in Panel B. B: CCF fluctuation due to photon noise, i.e., the difference between blue and green CCF in Panel A. When photon noise is small, CCF fluctuation due to photon noise is smaller than CCF fluctuation due to intrinsic CCF structures (blue CCF in Panel A). C: close-up for CCF peaks. D: close-up for CCF regions where we define fluctuation level of a CCF. We use the RMS of either the first quarter or the fourth quarter to calculate CCF fluctuation level. E: close-up to show CCFs of different cases. 
\label{fig:CCF_illustration}}
\end{figure*}

Once the detected spectra are obtained, we perform the data processing steps required to extract the planet signal using the cross correlation technique~\citep{Konopacky2013, Schwarz2016}. First, the sky emission spectrum is subtracted from the planet spectrum, and the planet spectrum is corrected for telluric absorption and stellar lines, which results in a so-called reduced spectrum. We note that since telluric removal is divisive and stellar removal is subtractive, stellar removal needs additional care in the  presence of significant planet light and abundant stellar lines \citep{Schwarz2016}. Then, the detected planet spectrum is cross-correlated with a synthetic planet template spectrum. For ground-based observations, the spectra used in the cross-correlation are high-pass filtered to remove the spectral continuum component ($<$100 cycles per micron). For space-based observations of planets whose spectra are dominated by reflected light, we remove the continuum by dividing the reflected light spectrum by the stellar spectrum. The cross-correlation between the reduced spectrum and the synthetic spectrum results in a CCF. The peak of the CCF is compared with the fluctuation level of the CCF (illustrated in Fig. \ref{fig:CCF_illustration}). We define CCF SNR as the ratio of CCF peak value and the RMS of CCF fluctuation. To calculate the RMS, we use either the first or the fourth quarter of CCF, whichever is further awary from the CCF peak. In order to be qualified as a significant detection, we require that (1) CCF SNR is higher than 3 and (2) the RV of CCF peak is consistent with the input planet RV within one resolution element. Any significant detection of the CCF peak is equivalent to detecting the planet. To detect a certain molecular species in the planet spectrum (e.g., CO, H$_2$O), we simply repeat the same process using a synthetic planet template spectrum consisting of only lines from that single molecular species. 

For ground-based observations, the detection of the CCF peak is hampered by the Earth's atmosphere. This is especially the case if the molecular species of interest is also present in the Earth's atmosphere, e.g., O$_2$, H$_2$O and CO$_2$. In such cases, the CCF peak could be caused by residuals from the removal of telluric absorption lines. To distinguish the origin of the CCF peak, we use the fact that the RV of an exoplanet changes by tens of km/s due to its orbital motion and the Earth's barycentric motion whereas the RV variation of telluric lines stays within tens of m/s. To measure an RV change of tens of km/s, the spectral resolution needs to be at least a few thousand at moderate SNR. Therefore, we consider only spectral resolutions higher than R=1,000 for ground-based observations. The ability of a spectrograph to distinguish the signal from an exoplanet and the signal from the Earth's atmosphere in RV space improves with increased spectral resolution. For space-based observations, the spectral resolution may be as low as R=25.

\section{Fiber Injection Unit, Upgraded NIRSPEC, and KPIC at Keck}
\label{sec:instrument}

While the framework described in \S \ref{sec:simulation} is a general-purpose pipeline to simulate performance of any HDC instruments, we will use the pipeline to study the prospect of the Keck Planet Imager and Characterizer~\citep[KPIC,][]{Mawet2016}, an HDC instrument that is being developed at Keck telescope.

KPIC is a four-pronged upgrade of the Keck adaptive optics facility. The first upgrade component is the addition of a high performance small inner working angle $L$-band vortex coronagraph to NIRC2~\citep{Absil2016}. This operation was successfully carried out in 2015 and is now available to the Keck community in shared risk mode. The upgrade not only came with a brand new coronagraph focal plane mask, but also a suite of software packages to automate the coronagraph acquisition procedure, including automatic ultra-precise centering~\citep{Huby2015}, speckle nulling wavefront control~\citep{Bottom2016}, and an open source python-based data reduction package~\citep{Gonzalez2016}. The second upgrade component is an infrared pyramid wavefront sensor demonstration, and potential facility for the Keck II adaptive optics system. The third upgrade component is a higher-order deformable mirror paired with the infrared pyramid sensor, followed by a new single-stage coronagraph. Finally, the fourth component of the KPIC is the fiber injection unit (FIU). 

The FIU is at the core of the KPIC instrument upgrade, which links the Keck II telescope AO bench to NIRSPEC, the current R$\sim$25,000 workhorse infrared spectrograph at Keck. The FIU focuses the light from a target of interest into single mode fibers after the AO system with minimal losses and the fiber outputs are reformatted to fit the slit plane of NIRSPEC.

In 2018, the UCLA IR lab will equip NIRSPEC with a new 5-$\mu$m cutoff, 2048x2048 pixel HgCdTe H2RG detector from Teledyne~\citep{Martin2014}. This new device offers reduced read noise and dark current, as well as improved cosmetics, superior flat-fielding, a modest improvement in quantum efficiency in $H$ and especially $K$ band, and the enhanced stability of modern electronics. Most critical for HRS is the H2RG’s smaller pixel scale of 18 $\mu$m (vs. 27 $\mu$m for the existing Aladdin device) which directly improves spectral resolution with the same grating arrangement from 25,000 to 37,500 with a 0$^{\prime\prime}$.29 slit and 3-pixel sampling.

Simulations in \S \ref{sec:ScienceCases} are based on the expected performance of KPIC at various stages of development. 

\section{Ground-based Observations of Directly-Imaged Planets with HDC}
\label{sec:ScienceCases}

\subsection{HR 8799 e}
\label{sec:HR8799e}

Planet e is the most challenging planet to observe among the 4 known planets in HR 8799 system because of its proximity to the host star ($\simeq 0^{\prime\prime}.4$). Following the methods detailed in \S \ref{sec:simulation}, we simulate observations with an HDC instrument (the FIU and upgraded NIRSPEC) at Keck. Input parameters for the planet, star, telescope, and instrument are provided in Table \ref{tab:telescope_instrument} and \ref{tab:HR8799e}.

BT-Settl model spectra are used as the input spectra. We use $T_{\rm{eff}} = 1200$~K and 7400 K and log(g)~=~3.5 and 4.5 for the planet and the star, respectively. The metallicity [Fe/H] is set to zero for both planet and star. 

\begin{figure*}
\epsscale{1.25}
\plotone{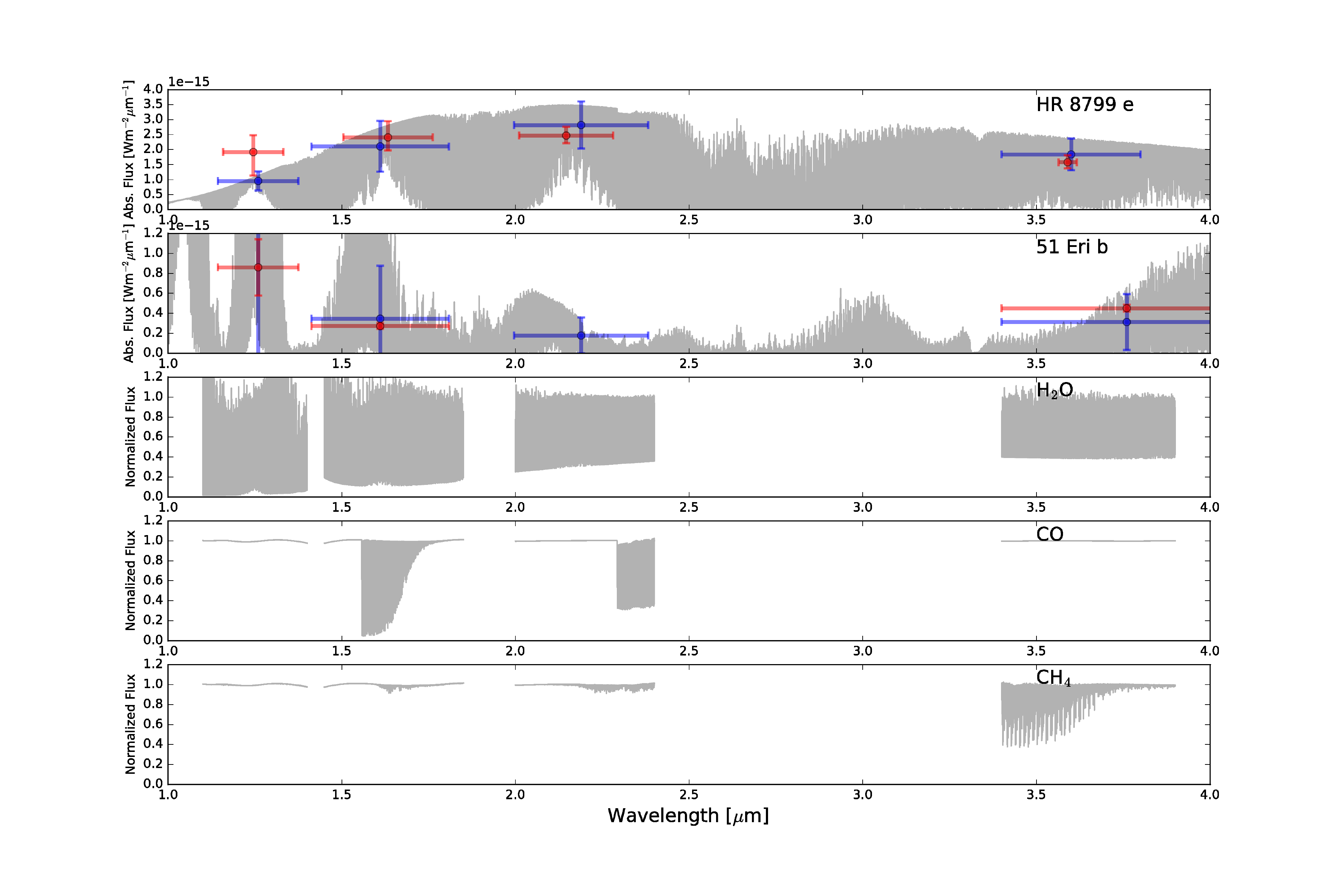}
\caption{Top two panels: BT-Settl spectra from HR 8799 e and 51 Eri b and comparison of absolute flux between model (blue) and observation~\citep[red, ][]{Bonnefoy2015, Macintosh2015} in different photometric bands. Bottom three panels: normalized spectra for individual molecular species. These spectra are used for detection of molecular species in the atmosphere of HR 8799 e and 51 Eri b.
\label{fig:Spec_HR8799_51Eri_mol_by_mol}}
\end{figure*}

The flux from the planet and star is adjusted such that the model flux is consistent with the absolute flux measured from photometry. We ensure that the adjusted flux matches with result from \citet{Bonnefoy2015} within uncertainties (see Fig.~\ref{fig:Spec_HR8799_51Eri_mol_by_mol}). 

We consider two cases: (1) we have perfect knowledge of the planet spectrum and (2) we have limited information about the intrinsic planet spectrum. In the first case, we use the BT-Settl model spectrum that is used to generate observations as the template. As a result, the input spectrum and the template spectrum are the same. In the second case, a combined molecule-by-molecule spectrum of CO, CH$_4$, and H$_2$O is used as the template. As a result, the input spectrum and the template spectrum are independently generated and may not necessarily the same. 

\subsubsection{Limiting Factors for CCF SNR}
\label{sec:limiting_factor}

\begin{figure*}
\epsscale{1.0}
\plotone{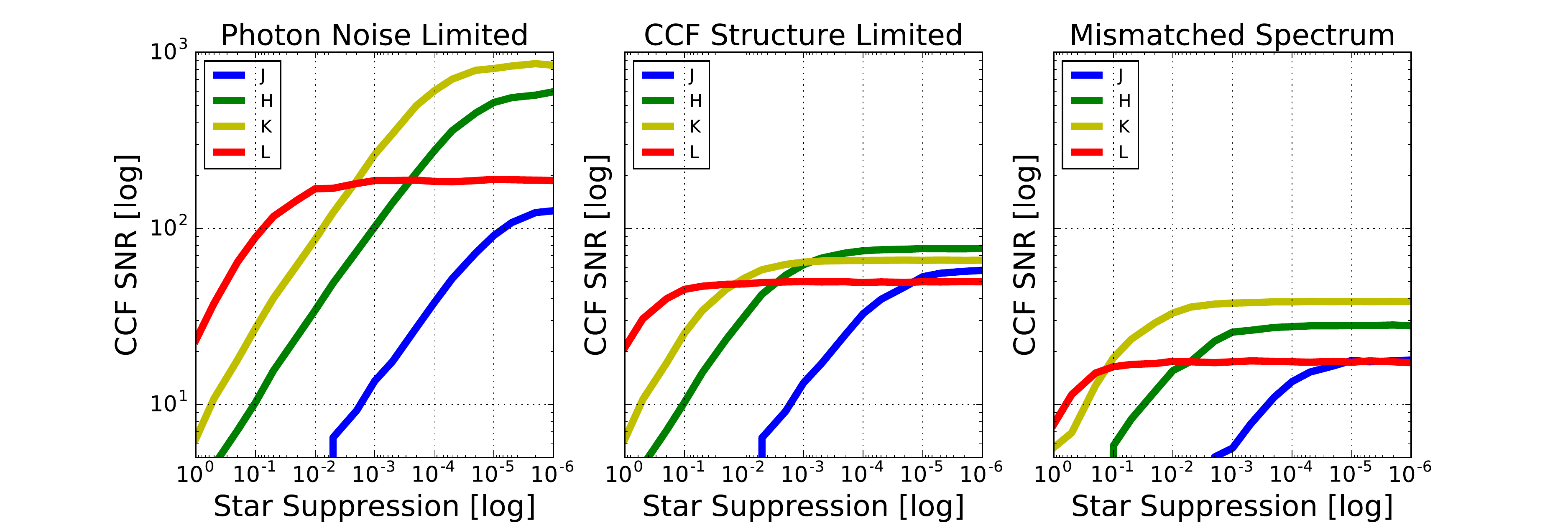}
\caption{CCF SNR vs. star light suppression level for HR 8799 e in 1-hr exposure time for three cases (see discussion in \S \ref{sec:limiting_factor}). Simulation parameters for the planet, star, telescope, and instrument are provided in Table \ref{tab:telescope_instrument} and \ref{tab:HR8799e}.
\label{fig:HR8799e}}
\end{figure*}

We simulate 100 observations at each star light suppression level and for each band. The median value of these simulations is reported in the following discussion. We consider three scenarios in the CCF SNR calculation (Fig.~\ref{fig:HR8799e}). In the {\bf{CCF structure limited}} case, the CCF SNR is limited by the intrinsic structure in regions where we calculate the noise level. We use the RMS of the first quarter or the forth quarter to calculate the noise level of CCF (see Fig. \ref{fig:CCF_illustration} for illustration). If the CCF peak is in the first half of CCF, then we use the forth quarter for RMS calculation. Otherwise, we use the first quarter for RMS calculation. The velocity span of CCF is half of the bandwidth times the speed of light, which is the result of Fourier transform that is used in CCF calculation. 

In theory, one can remove the intrinsic CCF structure by subtracting the noiseless CCF from the noisy CCF. The remaining noise level is due to photon noise (see Panel B in Fig. \ref{fig:CCF_illustration}), which is the {\bf{photon noise limited}} case. The limiting photon noise can be from various sources. At low level of star light suppression, the dominating noise source is always the photon noise from the star. At deeper star light suppression, the limiting photon noise can be sky background emission (e..g, $L^{\prime}$ band) or the planet itself (e.g., $J$, $H$, $K_S$ band). The photon noise limited case is the most optimistic case in which we have perfect knowledge of the planet and the star. 

In practice, however, we do not know the noiseless planet and star spectra a priori, so we do not know the noiseless CCF. Therefore, CCF SNR is almost certainly limited by systematics. In addition to the CCF structure limited case, we also consider one case in which systematics dominates the CCF SNR. In the {\bf{mismatched spectrum}} case, we consider a mismatch between the observed and the template planet spectrum. For the observed planet spectrum, we use a BT-Settl spectrum with $T_{\rm{eff}} = 1200$ K and log(g) = 3.5. For the template planet spectrum, we use the combined spectrum of CO, CH$_4$, and H$_2$O as shown in Fig.~\ref{fig:Spec_HR8799_51Eri_mol_by_mol}. This scenario yields the lowest CCF SNR because of the spectrum mismatch. 

Although this case can potentially result in a low CCF SNR, it represents an opportunity for atmosphere retrieval: a more probable molecular abundance ratio, P-T profiles may be determined by varying model parameters to maximize the CCF peak. It highlights the importance of planet spectrum modeling and a good understanding of the systematics associated with the cross correlation method.

The three limiting cases represent the different stages of spectral retrieval. From a reduced spectrum, a template (likely mismatched) is used in the cross correlation which results in a CCF peak, assuming the template resembles the planet spectrum in the reduced spectrum. The result of this stage is equivalent to the mismatched spectrum case. Then, the template spectrum is optimized in order to maximize the CCF peak. During this process, planet atmospheric properties are inferred, including composition, abundance ratio, cloud patchiness, chemical equilibrium, etc. If the optimization process is successful, the CCF SNR is limited by the CCF’s intrinsic structure. At this stage, an auto-correlation function is calculated from the optimized template spectrum and subtracted from the optimized CCF to remove intrinsic structures. After the subtraction, the data reduction and spectral retrieval can potentially reach the photon-noise limit. 

\subsubsection{Optimal Band For Planet Detection}

Fig.~\ref{fig:HR8799e} shows CCF SNRs at star light suppression levels up to $10^{-6}$. At a low level of star light suppression ($>10^{-2}$), the $L^{\prime}$ band outperforms other bands because the planet/star contrast is favorable (see Table \ref{tab:HR8799e}). However, the $L^{\prime}$ curves level off quickly as the star light suppression level increases because sky background becomes the dominant noise source. In this case, increasing star light suppression level does not improve the CCF SNR. However, we note that the starlight suppression at the beginning of the plateau depends on the brightness of a star. That is, deeper starlight suppression is needed to reach the background limit for brighter stars.

At deeper star light suppression, $H$ and $K_S$ band becomes the optimal bands that give the highest CCF SNR. The transition of performance between $L^{\prime}$ and $H$/$K_S$ band takes place at star light suppression levels between $\sim10^{-1}$-$10^{-3}$ depending on different cases. 

For a  given  angular  separation,  there  is  a trade-off between operating wavelength and wavefront  quality. For instance, the Strehl ratio  is  worse  at shorter wavelengths, but spatial resolution improves. Coronagraph performance is usually better with more beam widths ($\lambda$/D in angle) separating the star and planet. In  our  simulations,  we  scale  the nominal 10\% throughput with the Strehl ratio to account for better wavefront quality at  longer  wavelengths, which results both in better  coronagraph  performance  and  fiber coupling  efficiency.  We  do  not  directly include  the  benefit  of  higher resolution at shorter wavelengths because angular separations (in units of $\lambda$/D, see Table \ref{tab:HR8799e}) for HR 8799 e are much larger than the spatial resolution of KPIC.

\subsubsection{Sensitivity Gain in HDC Observation}
\label{sec:gain_factor}

Compared to ground-based HCI observations of HR 8799 e, HDC observations would provide a significant gain in sensitivity. In L$^\prime$ band, the detection significance is 5-10 for HCI only on Keck telescope~\citep{Currie2014}. In comparison, our simulations indicate that, at a level of star light suppression of $10^{-3}$, CCF SNR in L$^\prime$ is between $\sim$20 (mismatched spectrum case) and 200 (photon noise limited case). This is a factor of $\sim$2-40 gain in sensitivity with the help of HRS. The gain is because HRS serves as an additional filter for the planet signal. However, the gain in $L^\prime$ band is limited by strong sky emission. 

In other bands for which the sensitivity is not limited by the sky background but by the planet/star contrast, we expect an HDC instrument to provide an even higher gain in sensitivity. For example, the planet/star contrast for HR 8799 e is $\sim4\times10^{-5}$ in $K_S$, which may not be seen by an HCI instrument with star light suppression level of $10^{-3}$. With an HDC instrument, the planet can be detected with a CCF SNR of 40-250 (Fig.~\ref{fig:HR8799e}).

\subsubsection{Molecular Detection}

\begin{figure*}
\epsscale{1.25}
\plotone{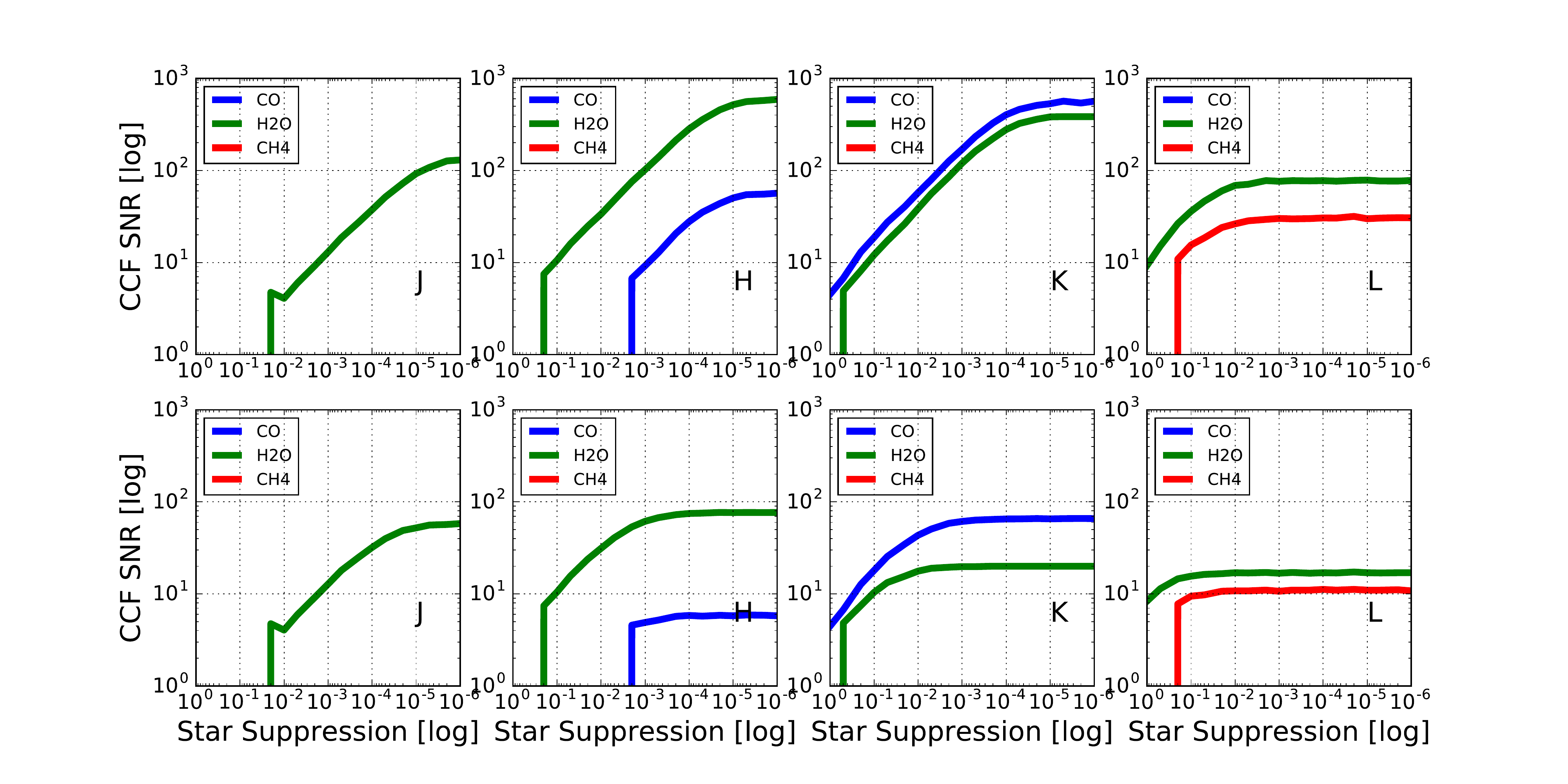}
\caption{CCF SNR for molecular detection in the atmosphere of HR 8799 e assuming 1-hr exposure time. Top rows are for the photon-noise limited case and bottom rows are for the mismatched spectrum case. Simulation parameters for the planet, star, telescope, and instrument are provided in Table \ref{tab:telescope_instrument} and \ref{tab:HR8799e}.
\label{fig:HR8799_mol_JHKL}}
\end{figure*}

In addition to planet detection using the cross correlation method, we also consider detecting individual molecular species in the atmosphere of a planet. The template spectrum for a single molecular species is generated as described in \S \ref{sec:simulation_spec}. In total, we generate spectra for three molecular species: CO, H$_2$O and CH$_4$, plotted in Fig.~\ref{fig:Spec_HR8799_51Eri_mol_by_mol} along with BT-Settl spectra for HR 8799 e and 51 Eri b. 

While $H$, $K_S$ and $L^{\prime}$ bands are identified as the optimal bands for HR 8799 e detection, we investigate the potential for using all four bands in searching for molecular species in the atmosphere of HR 8799 e. To do so, we cross correlate simulated observed planet spectrum with a template spectrum of an individual molecular species.

Fig.~\ref{fig:HR8799_mol_JHKL} shows the CCF SNR as a function of star light suppression level for CO, H$_2$O and CH$_4$ for $J$, $H$, $K_S$, and $L^{\prime}$ band observations. The optimal bands for CO, H$_2$O and CH$_4$ detection are $K_S$, $H$, and $L^{\prime}$, respectively. The differences between the optimal bands for planet detection and molecular species detection highlights the need for multi-band high-resolution spectroscopy.

When comparing to previous studies, our finding in $K_S$ band is consistent with Keck OSIRIS observations of HR 8799 c. Planet c has similar effective temperature and surface temperature HR 8799 e. With a star light suppression level of $\sim10^{-2}$, ~\citet{Konopacky2013} detected CO and H$_2$O in HR 8799 c with Keck OSIRIS at a CCF SNR of $\sim$10. The lower CCF SNR than what is predicted in Fig. \ref{fig:HR8799_mol_JHKL} can be attributed to lower spectral resolution and higher detector noise. 

The sharp drop of CCF SNR at low levels of starlight suppression in all subplots of Fig. \ref{fig:HR8799_mol_JHKL} is due to the criteria for planet/molecular detection in our simulation. In order to be qualified as a significant detection, we require that (1) CCF SNR is higher than 3 and (2) the RV of CCF peak is consistent with the input planet RV within one resolution element. Without the second criterion, there may be interlopers from random CCF fluctuation due to noise that may be misinterpreted as CCF peaks. In practice, measured CCF RVs should also follow a pattern that is consistent with the planet orbits. Therefore, if $>$50\% of the simulations result in an inconsistent RV, we assign a zero value to CCF SNR. The result implies that the minimum CCF SNR is $\sim$10 to confirm that the absorption/emission signal is indeed from the planet. 
 
\subsection{51 Eri b}
\label{sec:51Erib}

51 Eri b~\citep{Macintosh2015} is the only directly-imaged planet whose inferred mass is within the planet mass regime according to both cold-start and hot-start models~\citep{Bowler2016}. Furthermore, its brightness contrast and angular separation are representative of the practical detection limits of current ground-based high-contrast imagers. We therefore simulate observations of 51 Eri b with an HDC instrument to provide a point of comparison with the current state-of-the-art. 

Input parameters for the planet, host star, telescope, and instrument are provided in Table \ref{tab:telescope_instrument} and \ref{tab:51Erib}. We use input spectra with $T_{\rm{eff}} = 700$ K and log(g) = 3.5 for the planet and $T_{\rm{eff}} = 7400$ K and log(g) = 4.0 for the star. The metallicity [Fe/H] is set to zero for both planet and star. 

We adjust the planet and star flux such that the model flux and the absolute flux measured from photometry are consistent within uncertainties (Fig.~\ref{fig:Spec_HR8799_51Eri_mol_by_mol}). We adopt values from~\citet{Macintosh2015} for the absolute flux measurement. Similar to the cross correlation calculation presented for HR 8799 e, we use the BT-Settl spectrum as input to simulate observations. For template spectrum used for cross correlation, we either use the same spectrum as the input planet spectrum, or the combined molecule-by-molecule spectrum of CO, CH$_4$, and H$_2$O. 

\subsubsection{Optimal Band For Planet Detection}
\label{sec:51Erib_optimal}

\begin{figure*}
\epsscale{1.0}
\plotone{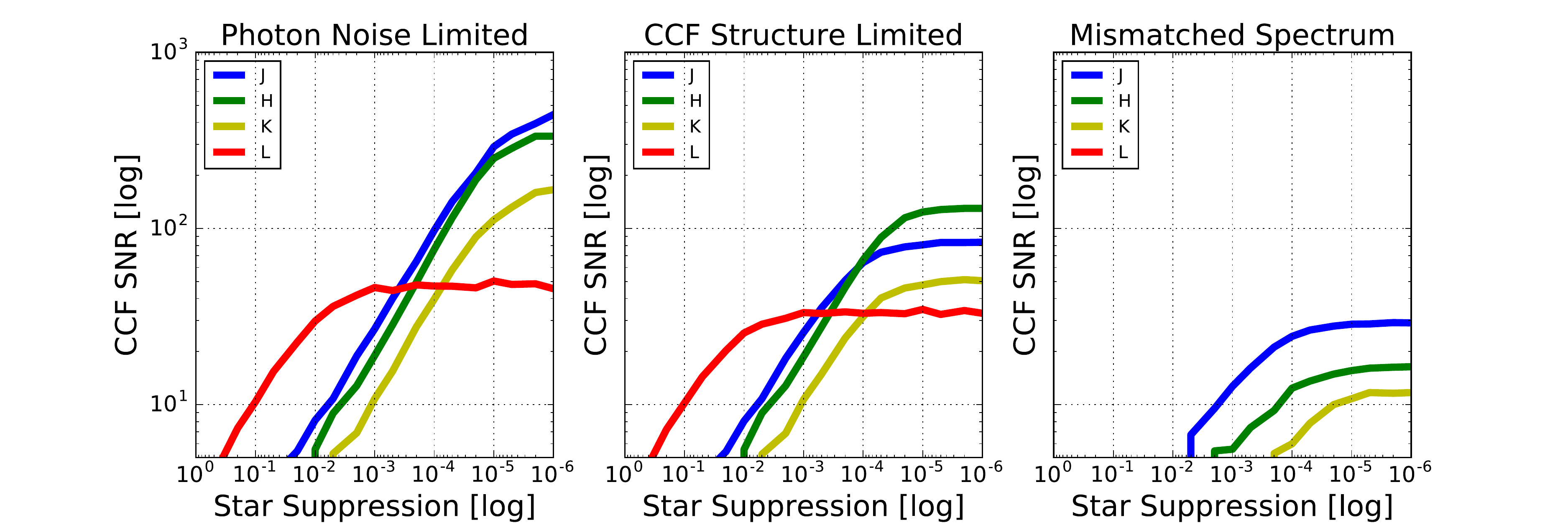}
\caption{CCF SNR vs. star light suppression level for 51 Eri b in 1-hr exposure time for three cases. Simulation parameters for the planet, star, telescope, and instrument are provided in Table \ref{tab:telescope_instrument} and \ref{tab:51Erib}. 
\label{fig:51Erib}}
\end{figure*}

Fig.~\ref{fig:51Erib} shows the CCF SNR for 51 Eri b in three cases. We again observe a decreasing trend of CCF SNR from the photon-noise limited case to the cases dominated by systematics. $J$ and $L^\prime$ bands are optimal bands for detecting 51 Eri b. $L^\prime$ band observation yields the highest CCF SNR for the photon-noise limited case and the CCF structure limited case, at low star light suppression levels ($>10^{-3}$). However, planet cannot be detected in $L^\prime$ band in the mismatched spectrum case. This is possibly because of a poor knowledge of $L^\prime$ band planet spectrum. $J$ band is the optimal band for the photon-noise limited case and the CCF structure limited case if star light suppression levels is better than a few times $10^{-4}$. In addition, $J$ band is also the optimal band for the mismatched spectrum case. This is largely due to the high photon flux from the planet in $J$ band. 

In the photon-noise limited case and the CCF structure limited case, we use planet template spectrum that is exactly the same as the planet spectrum used in simulating observation. This is to assume that we have full knowledge of the planet's spectrum. While this assumption leads to a much higher CCF SNR (see Fig.~\ref{fig:51Erib}), we cannot practically generate a perfect planet or molecular template spectrum. 

To demonstrate this point, we use the BT-Settl spectrum as an input to simulate the astrophysical signal. We use the combined molecular spectrum for CO, H$_2$O, and CH$_4$ as the template spectrum. As a result, CCF SNR is reduced for all bands (see Fig.~\ref{fig:51Erib}). Using an imperfect template in the cross correlation operation may even lead to missed detections of planets or particular molecular species. However, as mentioned in \S \ref{sec:limiting_factor}, the mismatched spectrum case also represents an opportunity for atmospheric retrieval.


\subsubsection{Molecular Detection}

\begin{figure*}
\epsscale{1.25}
\plotone{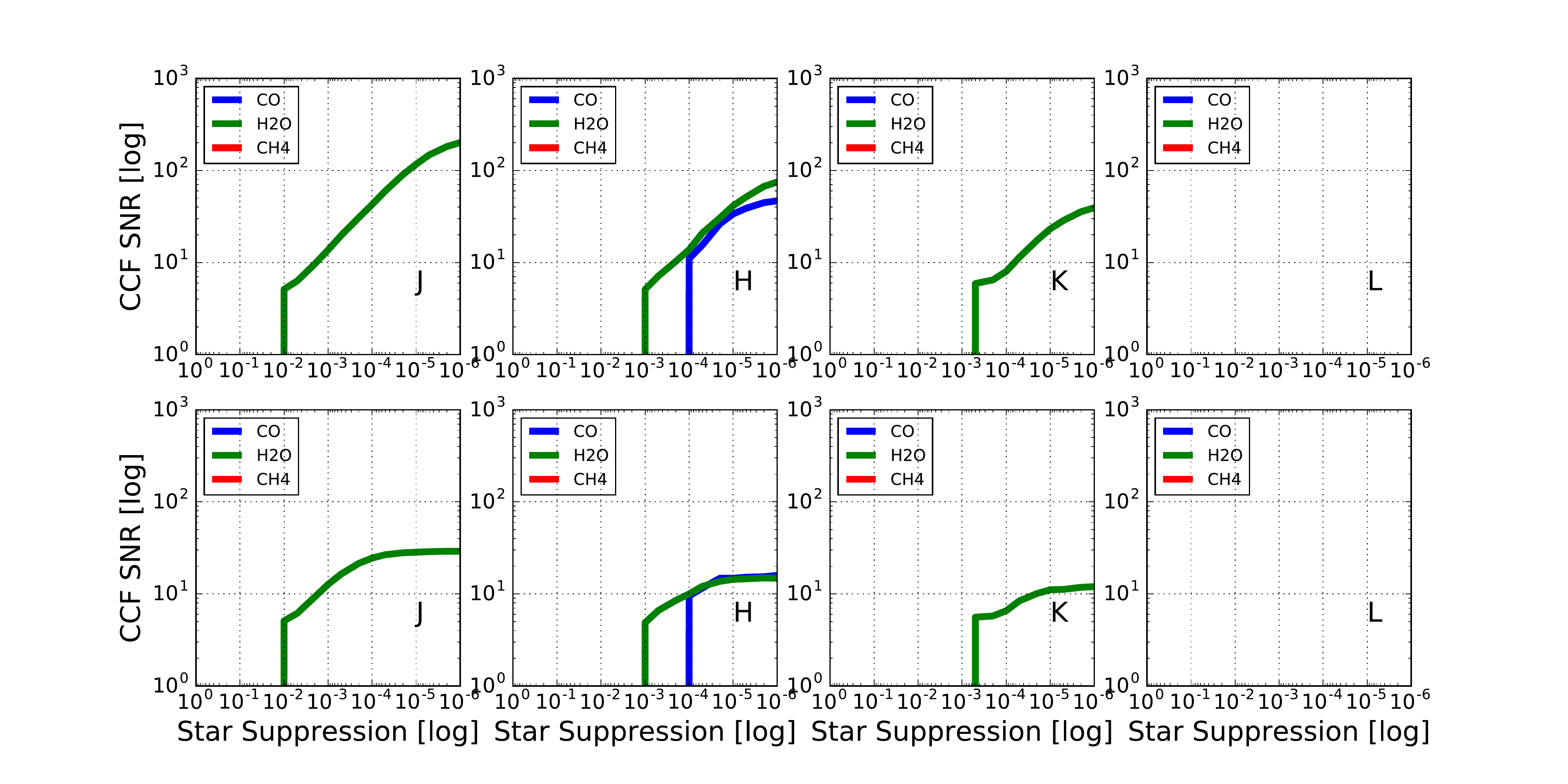}
\caption{CCF SNR for molecular detection in the atmosphere of 51 Eri b assuming 1-hr exposure time. Top rows are for the photon-noise limited case and bottom rows are for the mismatched spectrum case. Simulation parameters for the planet, star, telescope, and instrument are provided in Table \ref{tab:telescope_instrument} and \ref{tab:51Erib}.
\label{fig:51Erib_mol_JHKL}}
\end{figure*}

Fig.~\ref{fig:51Erib_mol_JHKL} shows the CCF SNR achieved by cross correlating the reduced spectrum with template spectrum of individual molecular species. Depending on the photon flux and the density and strength of the spectral lines, the optimal band is different for each species. H$_2$O is present in all $J$, $H$, $K_S$, and $L^\prime$ bands (see Fig.~\ref{fig:Spec_HR8799_51Eri_mol_by_mol}) and can be detected in $J$, $H$ and $K_S$ band. The highest CCF SNR is given in $J$ band. CO has lines in $H$ and $K_S$ band and can be detected in $H$ band. Although abundant CH$_4$ lines exist in $L^\prime$ band, CH$_4$ in 51 Eri b can not be detected due to much elevated sky background and much reduced photon flux compared to HR 8799 e.  

\section{Searching and Characterizing Earth-like Planets Around Low-Mass Stars with Ground-Based Extremely Large Telescopes}
\label{sec:Earths}

Searching for Earth-like planets and identifying molecular species in their atmospheres is one of the major science goals for ground-based extremely large telescopes and future space-based missions. Ground-based telescopes are generally larger than space-based telescopes and thus have the advantage of higher angular resolution at a given wavelength. On the other hand, space-based telescopes can achieve deeper star light suppression than ground-based instruments due to their vantage point outside our turbulent atmosphere. These differences in spatial resolution and achievable contrast levels affect the science objectives of space-based and ground-based missions for the study of Earth-like planets. Ground-based are more suitable in studying Earth-like planets around low-mass stars because of (1) less stringent requirements for star light suppression and (2) potentially improved inner working angle (IWA) due to the increased telescope aperture size. In comparison, space-based missions are better for targeting Earth-like planets around solar type stars because of (1) deeper star light suppression and (2) less stringent requirements for IWA. 

The recent discovery of Proxima Cen b~\citep{Escude2016} makes this Earth-like planet candidate an excellent target to characterize. However, this requires significantly upgraded capabilities of current telescopes~\citep{Lovis2016}. To demonstrate the potential of HDC for a 30-m class telescope,  we simulate observations of (1) Proxima Cen b and (2) an Earth-like planet in the habitable zone of a M dwarf at 5 pc. The second case represents a general case study whereas the first is the best case scenario owing to the proximity of Proxima Cen. 

We simulate observations in $J$, $H$, $K_S$ and $L^\prime$ band and find that the CCF SNR in $L^\prime$ does not reach the detection threshold within the considered star light suppression levels and spectral resolutions, so we only discuss $J$, $H$ and $K_S$ results here. While $L^\prime$ band is not an optimal band to search for planets in reflected light, longer wavelengths (e.g., $M$ and $N$ band) may be considered in the search for planet emission. Bandwidths for $J$ and $K_S$ bands are within 20\% and $H$ band is $\sim$25\%. While it is challenging to keep consistent star light suppression level over such a wide bandwidth, we consider the full wavelength range for these bands. In practice, suboptimal wide-band performance may be improved by multiple observations with narrower bandwidths.

\subsection{Simulation Setup}
\label{sec:mstar_planet_spec}

\begin{figure*}
\epsscale{1.18}
\plotone{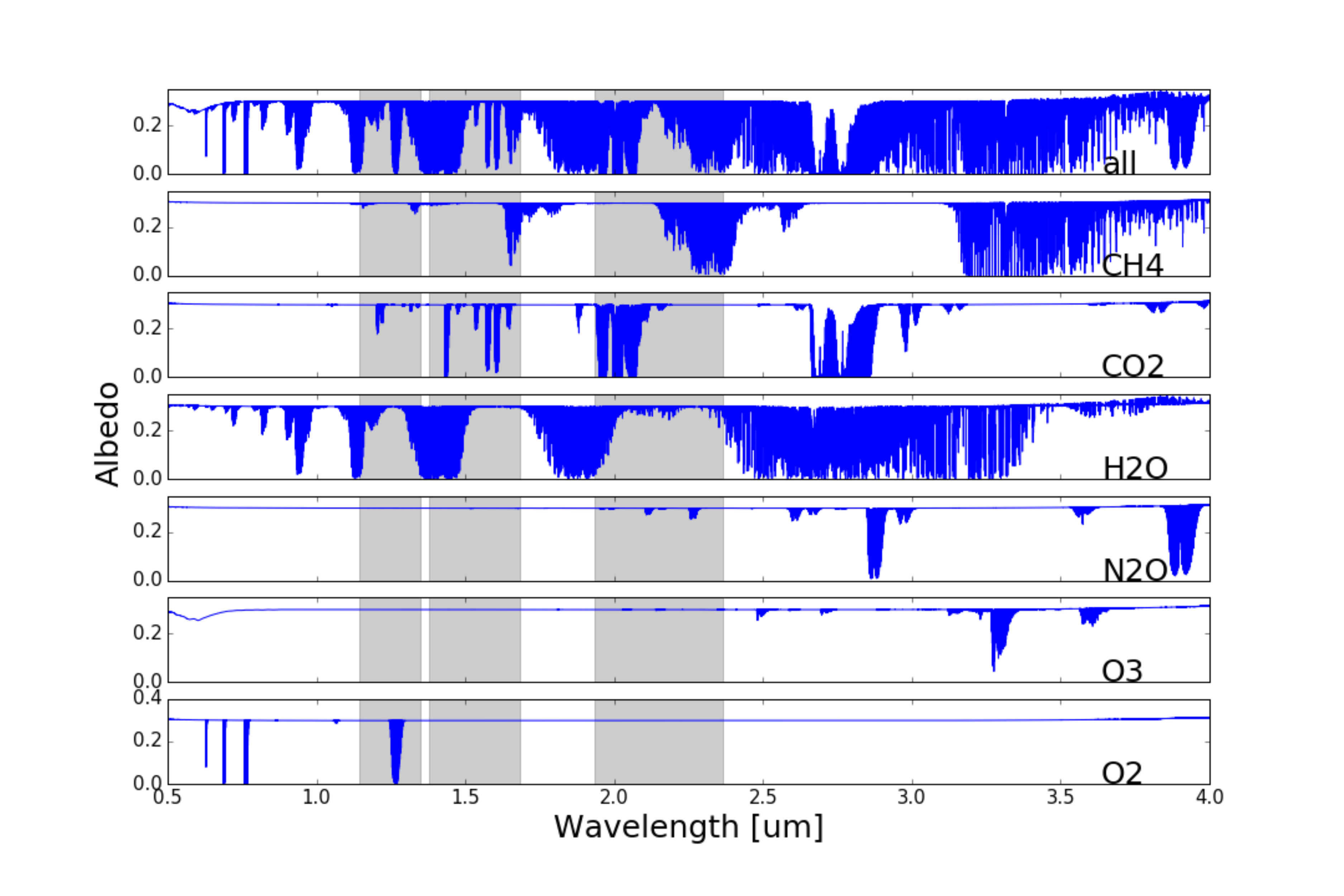}
\caption{Albedo spectrum of an Earth-like planet. We consider the average albedo between a high cloud case (high albedo) and cloud-free case (low albedo). Shaded regions are wavelength regions we consider to simulate observations for detecting molecular species with ground-based telescopes. For space-based observation, we consider a wavelength region from 0.5 to 1.7~$\mu$m.\label{fig:Molecular_absorption}}
\end{figure*}

We use the Earth albedo spectrum (Fig.~\ref{fig:Molecular_absorption}) for the planet, which is a product of stellar spectrum and the albedo spectrum. The absolute flux of the spectrum is then scaled with the planet radius, the planet-star separation, and the planet fractional illuminated area (i.e., phase function). We use the BT-Settl spectrum with $T_{\rm{eff}} = 3500$ K and log(g) = 4.5 as the input M dwarf spectrum. For Proxima Cen, we use the BT-Settl spectrum with $T_{\rm{eff}} = 3000$ K and log(g) = 5.0. The metallicity [Fe/H] is set to zero for all cases. The telescope and instrument parameters used in simulation can be found in Table~\ref{tab:Telescope_Instrument_Mdwarf_Earth}. Further information about the planet and star can be found in Tables \ref{tab:ProxCenb} and \ref{tab:Mdwarf_Earth}. 

\subsection{Results for $J$, $H$ and $K_S$ bands}
\label{sec:JHK}

\begin{figure*}
\epsscale{1.18}
\plotone{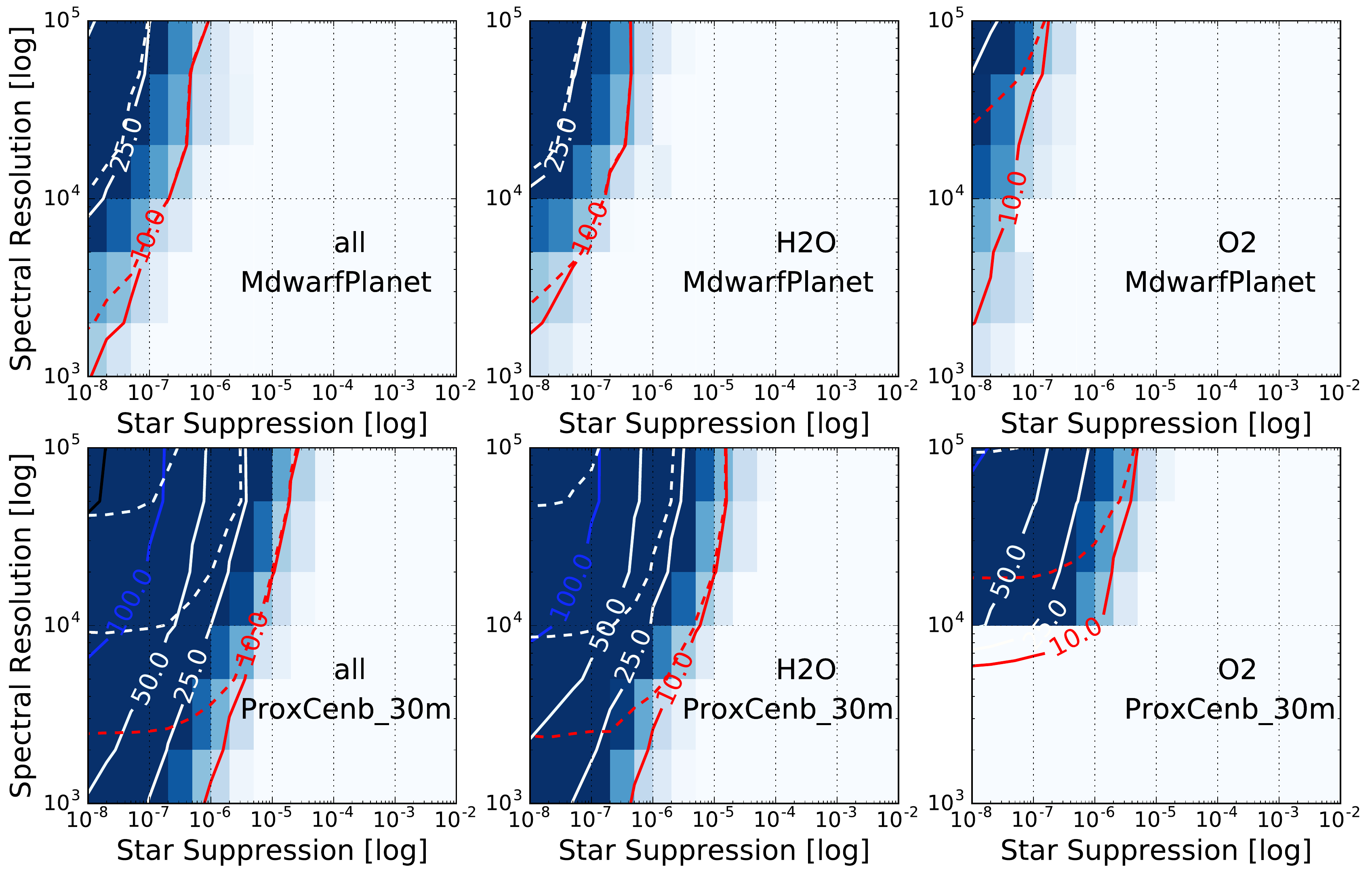}
\caption{CCF SNR contours for $J$-band simulation in phase space of spectral resolution and star light suppression level for different molecular species for three cases: (1) a 30-m telescope on a Earth-like planet around a M dwarf at 5 pc (top rows); and (2) a 30-m telescope on Proxima Cen b (bottom rows). Solid contours are for the photon-noise limited case and dashed contours are for the CCF structure limited case. Each panel is marked with the name of a molecular species indicating only lines of a given molecular species are used in cross correlation. ``All" means all lines are used. \label{fig:Ground_J}}
\end{figure*}

\begin{figure*}
\epsscale{1.18}
\plotone{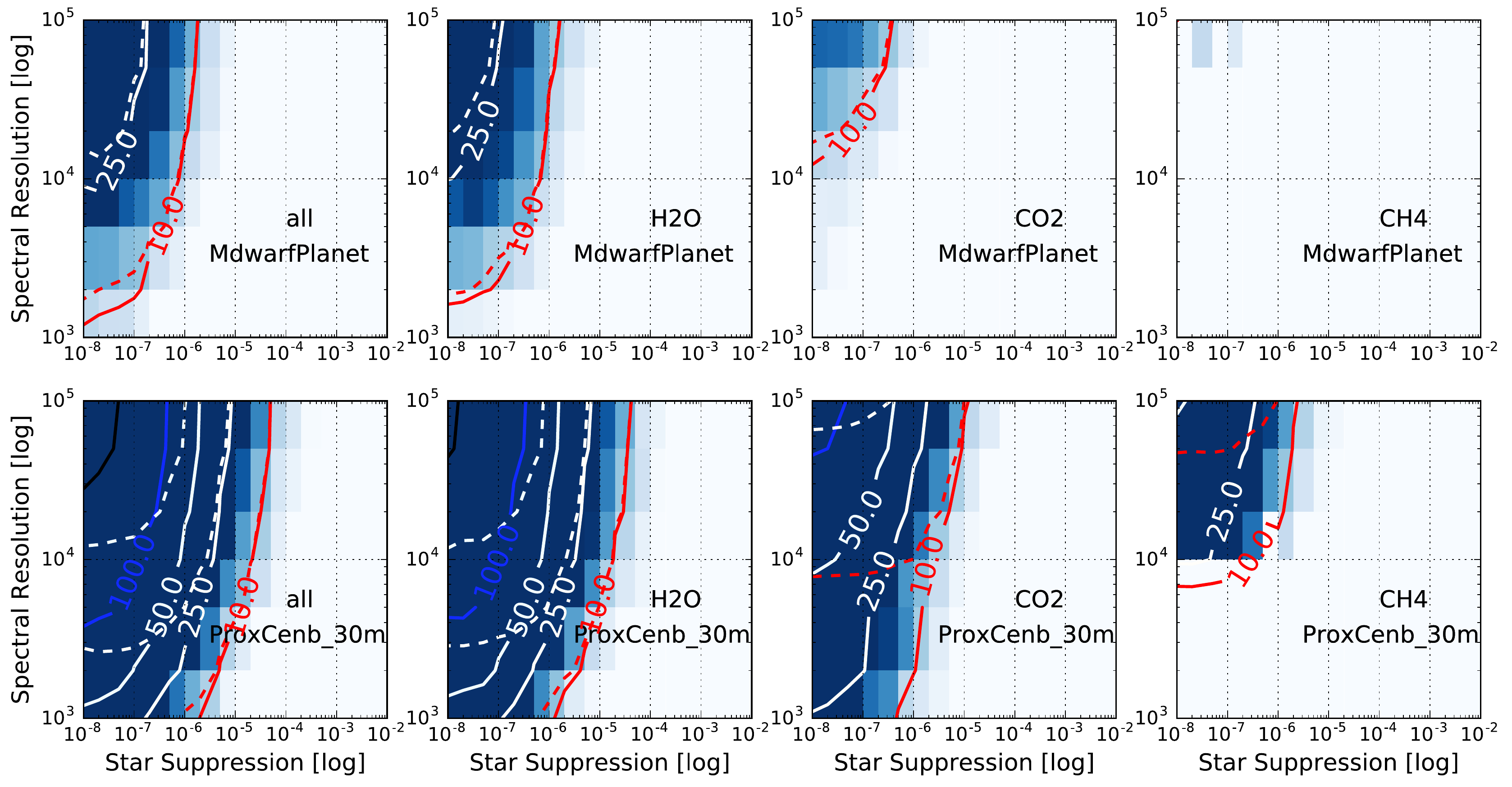}
\caption{Same as Fig.~\ref{fig:Ground_J} but for $H$-band simulation. \label{fig:Ground_H}}
\end{figure*}

\begin{figure*}
\epsscale{1.18}
\plotone{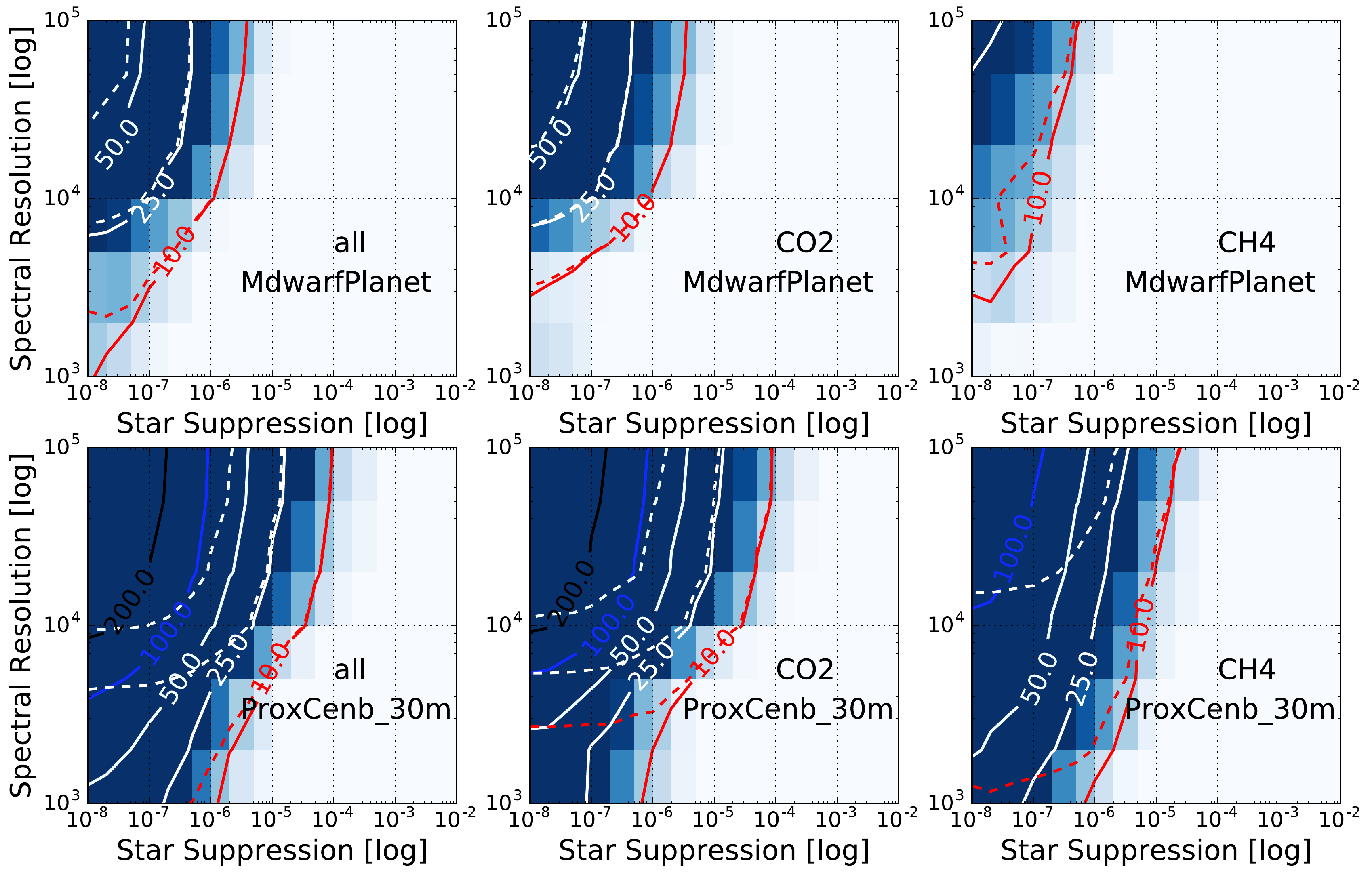}
\caption{Same as Fig.~\ref{fig:Ground_J} but for $K_S$-band simulation. \label{fig:Ground_K}}
\end{figure*}

\subsubsection{An Interplay Between HCI and HRS}

Fig.~\ref{fig:Ground_J}-\ref{fig:Ground_K} show the CCF SNR contours as a function of spectral resolution and star light suppression level for $J$, $H$, and $K_S$-band observations.  
The general trend is that the CCF SNR increases with higher spectral resolution and deeper levels of star light suppression. As a result, high spectral resolution relaxes the star light suppression requirements by orders of magnitude. This has significant implications for HDC observations: insufficient star light suppression may be compensated by increasing spectral resolution. 

The planet/star contrast is $\sim$10$^{-8}$-10$^{-7}$ for the M dwarf planet and Proxima Cen b systems. However, it is extremely challenging to achieve a star light suppression level of $\sim$10$^{-8}$ from the ground. With the help of HRS, the star light suppression requirement can be relaxed by about 2-3 orders of magnitude. While there is no clear pathway to achieve $\sim$10$^{-8}$ star light suppression levels with ground-based telescopes, 10$^{-5}$-10$^{-6}$ is a much more attainable goal, which is within reach of mainstream extreme AO systems currently operating on most 8-m - 10-m class telescopes.


\subsubsection{Star Light Suppression vs. Planet Signal}
\label{sec:suppression_signal}

Having a larger telescope aperture not only improves angular resolution, but is also critical for gathering sufficient signal. The improved signal increases the CCF SNR, thereby relaxing the requirements for star light suppression. Likewise, star light suppression requirements may be further relaxed by increasing signal via longer exposure times or improving instrument throughput. In all cases, the boost in sensitivity provided by an HDC instrument depends on how much signal the instrument receives and how it compares with the relevant noise sources.

\begin{figure}
\epsscale{1.2}
\plotone{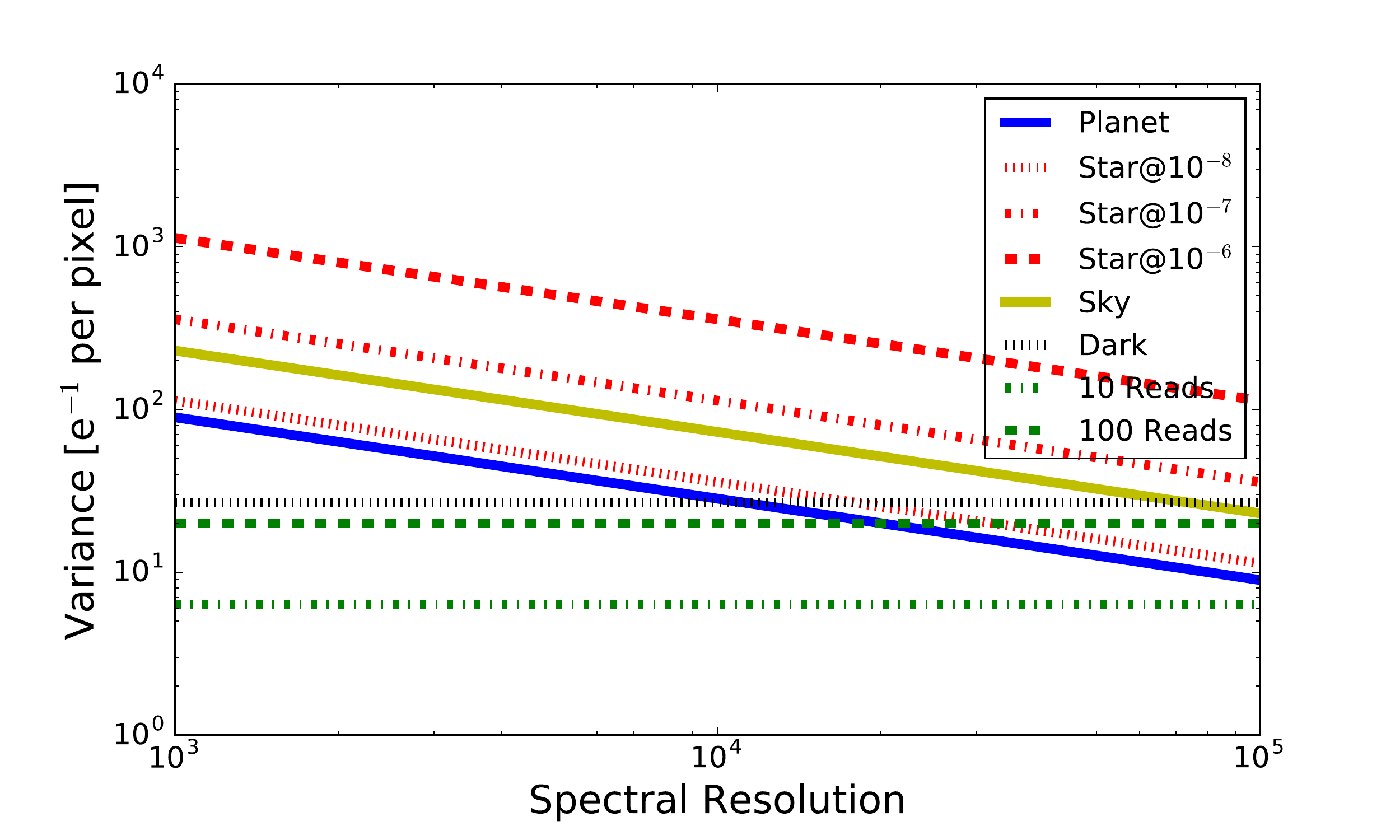}
\caption{A comparison of noise sources at different spectral resolutions for the case of 30-m telescope on a M dwarf planet in $K_S$ band.  
\label{fig:Ground_SNR}}
\end{figure}

Figure~\ref{fig:Ground_SNR} shows the noise sources for the case of a 30~m telescope observing an M dwarf planet in the $K_S$ band. The plot compares noise sources as a function of spectral resolution. Depending on the spectral resolution and star light suppression level, the dominating noise source can be sky background, photon noise and detector noise. For example, sky background noise dominates at deep star light suppression levels and low spectral resolution. Photon noise from the leaked stellar light dominates at low level of star light suppression. Detector noise dominates at deep star light suppression and high spectral resolution. 

Ground-based HDC instruments usually operate at high spectral resolution (R$\sim$100,000) and a star light suppression level that is a few orders of magnitude worse than the planet/star contrast. Therefore, the photon noise from the leaked star light is typically the dominant noise source. The detector and sky background noise sources are $\sim$5 times lower than the photon noise source at a $10^{-6}$ star light suppression level. The SNR per pixel is on the order of unity or less in the high spectral resolution regime. In order to achieve higher sensitivity at lower levels of star light suppression, increasing signal from the planet is the key. 

\subsubsection{Photon-Noise Limited vs. CCF Structure Limited}

The solid and dashed contours in Figs.~\ref{fig:Ground_J}-\ref{fig:Ground_K} represent two different cases: the photon noise limited case and the CCF structure limited case. In the high spectral resolution regime, the two contours usually agree with each other. The agreement between two sets of contours can be explained by the low SNR per pixel at high spectral resolution regime. In the situation of low SNR per pixel, CCF fluctuation is mainly due to photon noise. Therefore, CCF SNR in the the photon noise limited case is essentially the same case as the CCF structure limited case. In contrast, in the low spectral resolution regime, the SNR per pixel is higher as shown by Fig.~\ref{fig:Ground_SNR}. The CCF fluctuation is no longer due to photon noise, but due to intrinsic CCF structures. Therefore, higher SNR per pixel causes the CCF SNR in the two cases to deviate from each other. This is true for the simulations in all bands. We observe similar deviations in our simulations for HR 8799 e (\S \ref{sec:HR8799e}) and 51 Eri b (\S \ref{sec:51Erib}). 

\subsubsection{Molecular Detection}

If Earth-sized planets around M dwarfs have atmospheres similar to the Earth's, H$_2$O, O$_2$, CO$_2$ and CH$_4$ may potentially be detected with HDC instruments on 30-m class telescopes. On the other hand, N$_2$O and O$_3$ are not detectable because of their lack of lines in the considered wavelength range. O$_2$ can only be detected in the $J$ band, H$_2$O is detectable in the $J$ and $H$ bands. Searching for CO$_2$ and CH$_4$ is better conducted in the $K_S$ band because of the molecule line density and depth. For the M dwarf planet case, CH$_4$ cannot be detected in the $H$ band with a CCF SNR over 10. 

\section{Searching and Characterizing Earth-like Planets Around Solar-Type Stars with Space-based Telescopes}
\label{sec:sun_earth}

Space-based instruments may achieve deep star light suppression ($<10^{-8}$) and will, therefore, allow observations that would be extremely challenging from the ground; i.e., observing a Sun-Earth system for which the planet/star contrast is $\sim10^{-10}$. In addition, space-based observations are free from contamination due to the Earth's atmosphere that may cause confusion when detecting molecular species that exists in both the Earth's and exoplanet's atmosphere. However, telescope apertures for space-based observations are typically much smaller than ground-based facilities. Next-generation space-based telescopes for high-contrast observations will range from 4 to 16~m in diameter. The angular separation of a Sun-Earth system at 5 pc (0$^{\prime\prime}$.2) would be 2.4-7.3 $\lambda/$D in $H$ band and 5.2-15.5 $\lambda/$D in $r$ band, which is within the working angle range of the most coronagraphs. Starshades could have an inner working angle as small as 1 $\lambda/$D. However, probing more distant systems would (1) reduce the absolute flux from the planet and (2) potentially make the angular separation fall below the inner working angle. 

Future space-based exoplanet missions will likely to be limited to $H$ band and shorter wavelengths. Beyond $H$ band, the thermal background rises, requiring a cryogenic telescope and instruments, which significantly increases the cost of the mission. For our simulations of space telescopes, we consider a wavelength range covering 0.5 to 1.7~$\mu$m. Such a wide passband poses a challenge for wavefront control, which typically operates at a bandwidth of 10\% to 20\% \citep{Trauger2007}. In order to reach the full 0.5 to 1.7~$\mu$m wavelength coverage, multiple observations or simultaneous wavefront control/coronagraph channels are necessary. While these practical concerns are neglected in our simulations, we note that lines for molecular species such as O$_2$ and CO$_2$ concentrate in smaller bands. Characterizing these molecular species may only require tailoring the instrument channels to regions of the spectrum where these specific lines are abundant.

\subsection{Simulation Setup}
\label{sec:sun_earth_spec}
 
We use the same Earth albedo spectrum to generate the Earth-like planet's reflection spectrum as described in \S \ref{sec:mstar_planet_spec}. The BT-Settl spectrum with $T_{\rm{eff}} = 5800$ K and log(g) = 4.5 is used as the input solar type star spectrum. The metallicity [Fe/H] is set to zero. The telescope and instrument parameters used are listed in Table \ref{tab:Telescope_Instrument_Sun_Earth} and the planet and host star information can be found in Table \ref{tab:Sun_Earth}.

\subsection{Masked Cross Correlation}
\label{sec:mask_ccf}

Space-based observations offer an opportunity to detect molecular absorption bands at low spectral resolution for a large wavelength range without confusion by the Earth's atmosphere. The opportunity also comes with challenges for the cross correlation technique. First, O$_2$ and CO$_2$ have only a few narrow absorption bands over a wide wavelength range. Using the entire wavelength range would not increase CCF SNR, but instead introduce noise in the CCF. For example, cross correlating the observed spectrum with and O$_2$ template at low spectral resolution results in a higher peak at the H$_2$O absorption band than at O$_2$ bands simply because the H$_2$O band is deeper.

For ground-based observations, all CCF definitions are observables. The CCF peak is the highest value of CCF and the CCF noise is the RMS of certain parts of the CCF. While we prefer such definitions, we have to make a few adjustments in the CCF SNR calculations for the space-based case. First, we apply a masked cross correlation method to only select wavelength regions with absorption lines~\citep{Queloz1995, Baranne1996}. This approach alleviates the confusion of the regular cross correlation at low spectral resolutions. At a given spectral resolution, the cross correlation mask selects wavelength regions with absorption lines/bands deeper than 1\% and set the rest of the reduced spectrum to the median value. Here, we consider a wider range of spectral resolutions from R = 25 to R = 102,400. 

In principle, one can calculate the CCF in smaller wavelength blocks and then add the CCFs of different blocks with weights based on information content and SNR. However, this is impractical at low spectral resolution where there are fewer than 100 CCF data points across the wavelength range from 0.5 to 1.7 $\mu$m. With fewer than 10 data points (i.e., 10 spectral block divisions), it is impossible to get a meaningful statistical peak and noise level. At higher spectral resolution, calculating the CCF over a broad range of wavelengths is essentially the same as calculating CCFs over smaller wavelength blocks followed by co-adding the CCFs. In order to maintain a consistent treatment across all considered spectral resolutions, we calculate the CCF over the entire wavelength range from 0.5 to 1.7 $\mu$m.

\subsection{A New Definition of CCF SNR}
\label{sec:new_ccf_snr}

\begin{figure*}
\epsscale{1.18}
\plotone{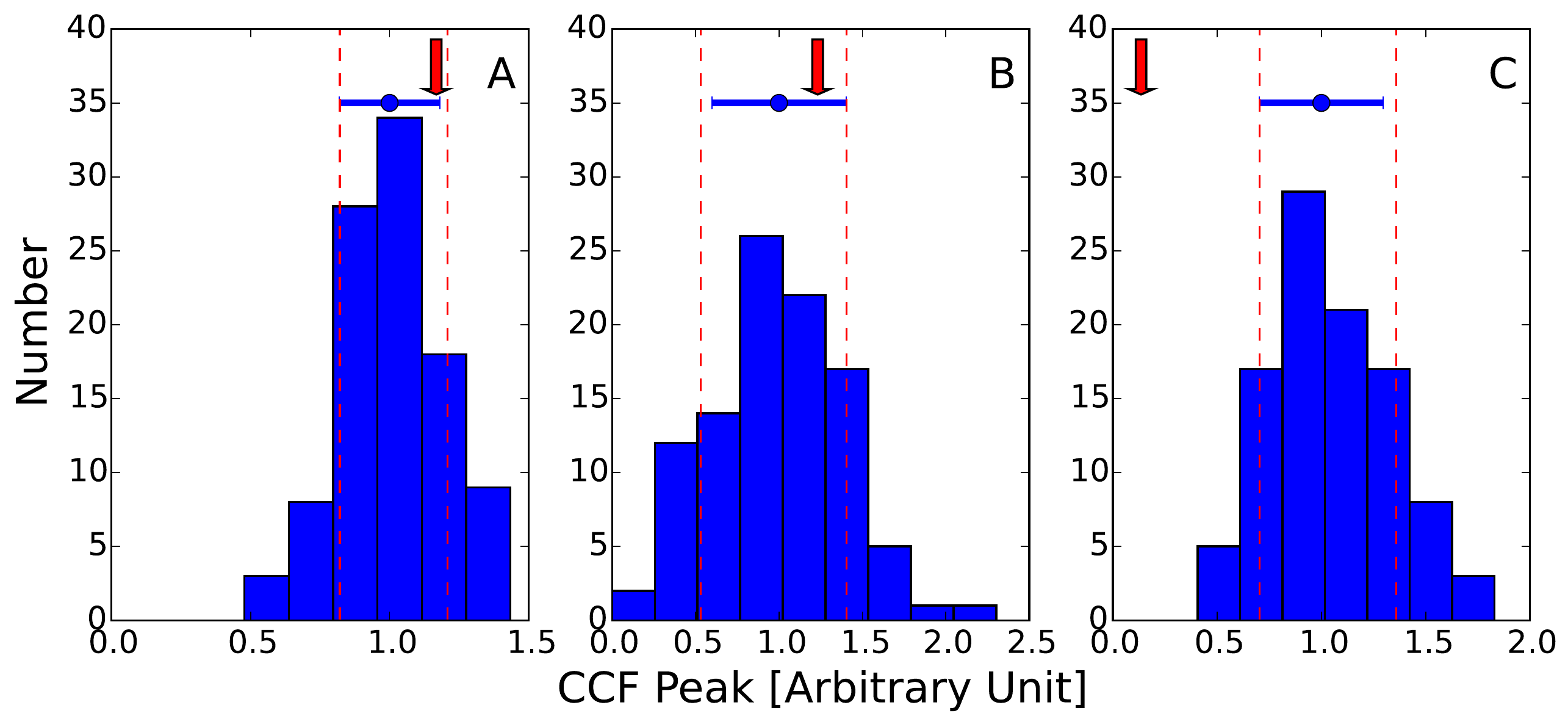}
\caption{Distribution of simulated CCF peaks. Red arrows are values of CCF Peaks in the noiseless case. Blue dots and errorbars are median values and standard deviations of the distributions. Red dashed lines mark the first and last 15\% percentile of distribution. Panel A: the distribution of CCF peaks is well separated from zero and the median is consistent with noiseless CCF peak value. Panel B: the distribution of CCF peaks is not well separated from zero because the lowest-valued bin has a non-zero number. Panel C: the distribution of CCF peaks is well separated from zero but the median is not consistent with noiseless CCF peak value. CCF peaks in this case are caused by random fluctuation induced by noise. \label{fig:hist_illustration}}
\end{figure*}

When using the masked cross correlation technique over a broad wavelength range, only parts of the CCF carries signal whereas the rest is flat. As a result, the previous definition of structure limited CCF SNR does not apply. In addition, the previous definition of photon-noise limited CCF SNR does not apply either. This is because the photon noise can be different by a factor of a few times from one end of the spectrum to the other. Although one could use the average of two ends of a spectrum to calculate the photon-noise limited CCF SNR, we choose the following way of defining the CCF SNR for the broadband wavelength coverage case. 

We simulate 100 observations, record all the CCF peak values, and make a histogram of the CCF peaks distribution divided into 10 bins from zero to the maximum CCF peak (as shown in Fig. \ref{fig:hist_illustration}). The CCF SNR is defined as the ratio between the median of simulated CCF peaks distribution and the standard deviation of the CCF peak distribution because the histogram is a reasonable approximation to a Gaussian distribution. As a result, a significant CCF peak should have a distribution that is well separated from zero (Panel A in Fig. \ref{fig:hist_illustration}), i.e., the lowest-valued bin that includes zero should have no data point from 100 simulations. This is roughly equivalent to a 3-$\sigma$ limit because there are 100 simulated peaks and none of them are consistent with zero (p$<$1\%). For a peak that is not well separated from zero (Panel B in Fig. \ref{fig:hist_illustration}), we mark the corresponding CCF SNR as zero. This happens when the height of the lowest-valued bin is not zero. 

If the CCF peak is caused by random variations rather than the planet signal (Panel C in Fig. \ref{fig:hist_illustration}), the CCF peak may be significant and well separated from zero. However, the center of the distribution of CCF peaks should be separated from the noiseless CCF peak. Therefore, we mark the CCF SNR as zero if the noiseless CCF peak is in the first 15\% or the last 15\% percentile of simulated CCF peak distribution. We choose this percentile for the following reasons. First, the boundaries are roughly consistent with 1-$\sigma$ range of simulated CCF peaks (red dashed lines vs. blue error bars). This ensures that inferred CCF peak is consistent with noiseless CCF peak within 1-$\sigma$. Second, the first 15\% percentile also roughly marks the 1-$\sigma$ lower boundary of simulated CCF peak distribution. This helps to exclude significant CCF peaks due to elevated noise levels with 1-$\sigma$ significance (Panel C). A higher-valued percentile cut would shrink the spacing between two dashed lines and therefore exclude more true positives (e.g., Panel A). On the other hand, a lower-valued percentile cut would include more false positives caused by random noise (e.g., Panel C). 

The CCF SNR definition used in this section requires multiple iterations to get a distribution of CCF peaks and to infer the significance of the CCF peak. In practice, only one CCF is obtained for one observation. In the presence of random systematics such as speckle chromatic noise at low spectral resolution, it is difficult to assess whether an observed CCF peak is caused by planet signal or due to random systematics. Simulations that incorporate our best knowledge of noise sources and systematics may be the only solution to quantify detection significance.  

At low spectral resolutions, the cross correlation may not be the most optimal way to detect planets/molecules. A more straightforward way is to conduct a conventional ADI/SDI sequence, detecting the planet, obtaining a low-resolution spectrum, and inferring molecular presence by measuring absorption band depth. However, in order to compare instrument performance over a broad range of spectral resolutions, we are compelled to apply the same cross correlation technique at all spectral resolutions for consistency and comparison purposes. 

\subsection{Speckle Noise and Its Chromaticity}
\label{sec:detnoise}

The spectral signature of speckles at deep star light suppression may resemble broad absorption bands, which mimic features in the planet's spectrum~\citep{Krist2008}. These artifacts are caused by wavelength dependent wavefront errors after amplitude and phase correction using a wavefront control scheme, such as electric field conjugation (EFC) \citep{Trauger2007,Groff2016}. 

\begin{figure}
\centering
\includegraphics[width=\linewidth]{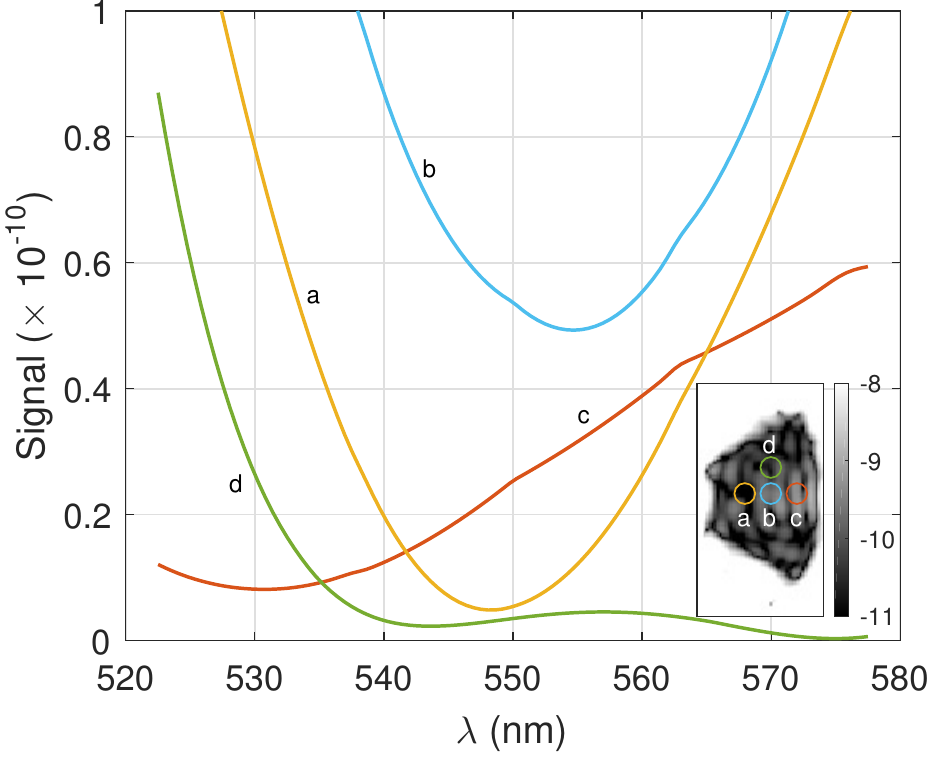}
\caption{High-resolution spectral signature of stellar speckles in a single mode fiber fed spectrograph. (inset)~Simulated irradiance in the dark hole, on a log scale, at $\lambda_0=550$~nm. The circles indicate the locations of the representative samples shown. All values are normalized to the peak of the stellar PSF prior to the coronagraph. 
\label{fig:Speckle_spectra}}
\end{figure}

To determine the HRS signature of speckles, we simulated a notional space telescope with realistic optical surface errors and a coronagraph instrument with two deformable mirrors, each with 16 actuators across the beam diameter. A dark hole was generated in the residual star light (10\% passband about $\lambda_0=550$ nm) within a 60$^\circ$ wedge-shaped region extending from 3-10~$\lambda/D$ using EFC (see Fig. \ref{fig:Speckle_spectra}, inset). The fiber coupling efficiency of the stellar field was calculated assuming a single mode fiber with a fundamental mode diameter of $\lambda/D$. Figure \ref{fig:Speckle_spectra} shows the estimated stellar signal detected by the spectrograph at example locations within the dark hole, indicated by the color circles in the inset. We find that the spectra of speckles generally take the form of low order polynomials. Since wavefront control simulations tend to be computationally intensive, we approximate this effect by generating low order spline function with points anchored at the edge of wavelength range for wavefront control and at an optimized wavelength. 

The wavelength coverage of our HDC observation simulations is 0.5 to 1.7 $\mu$m. However, it is not possible to use the deformable mirrors to generate a dark hole over such a wide range, so it is instead assumed that the wavefront control is performed at 0.1 $\mu$m intervals. For each reduced planet spectrum, we inject randomly generated speckle chromatic noise to study its impact on the cross correlation technique. To remove the low frequency speckle chromatic noise, we apply a high-pass filter before cross correlating with template spectrum. Some of the intrinsic planet absorption may also be removed since the absorption bands and the speckles have similar frequency content in the spectral domain.

\subsection{Simulating LUVOIR Observations}

The large ultraviolet, optical and infrared telescope (LUVOIR) is a concept study for a large next-generation space telescope \citep{Crooke2016}. The size of LUVOIR is not defined yet, but will likely be in the 10 to 16 meter range. Here, we conservatively select 12 meter for our simulations. The study of exoplanets will be one of its major scientific objectives. 

Unlike the ground-based simulations for HR 8799 e and 51 Eri b, we consider the photon-noise limited case. Detector noise (both readout noise and dark current) is set to zero. The availability of zero-noise detectors for space-based coronagraphic missions has been a concern and recently identified as a technology gap. It is now subject to growing awareness and research~\citep{Rauscher2016}. The impact of detector noise and speckle chromatic noise will also be discussed later in this section.

We note that the CCF structure limited case and the mismatch spectrum case are not considered for space-based observation. First, the CCF structure limited case is unlikely the case for Earth-like planet observations which are usually in low SNR regime. Second, there is no mismatch spectrum case since we use the same albedo spectrum for the input planet spectrum and the template spectrum for cross correlation. Therefore, the results shown below should be interpreted as an optimistic prediction of the performance for upcoming space-based instruments and missions. 

\subsubsection{Planet and Molecular Detection}
\begin{figure*}
\epsscale{1.2}
\plotone{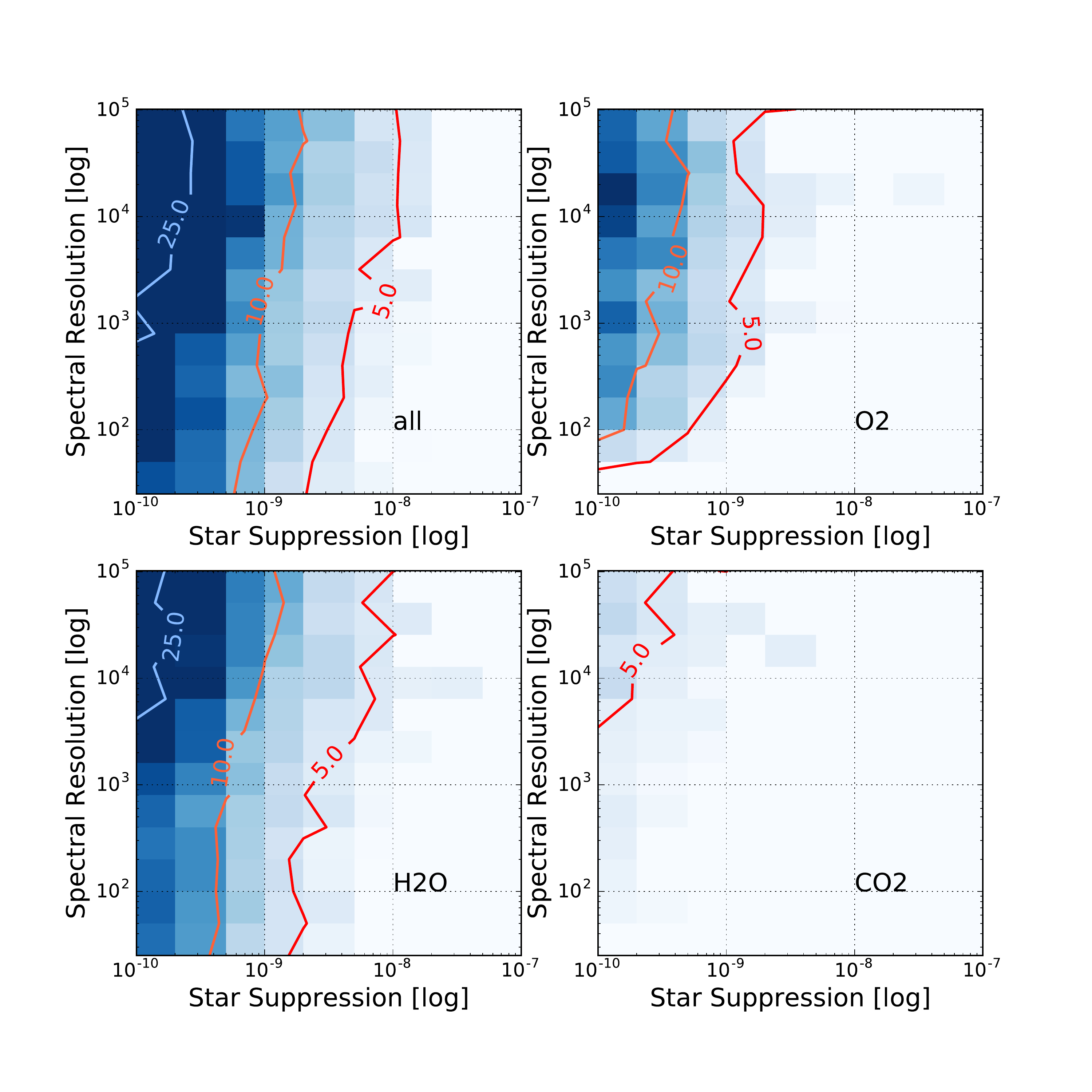}
\caption{CCF SNR contours for LUVOIR simulation in phase space of spectral resolution and star light suppression level for different molecular species. Detector noise (readout noise and dark current) is assumed to be zero. No speckle chromatic noise is considered. 
\label{fig:Space_molecular_detection}}
\end{figure*}

\begin{figure*}
\epsscale{1.2}
\plotone{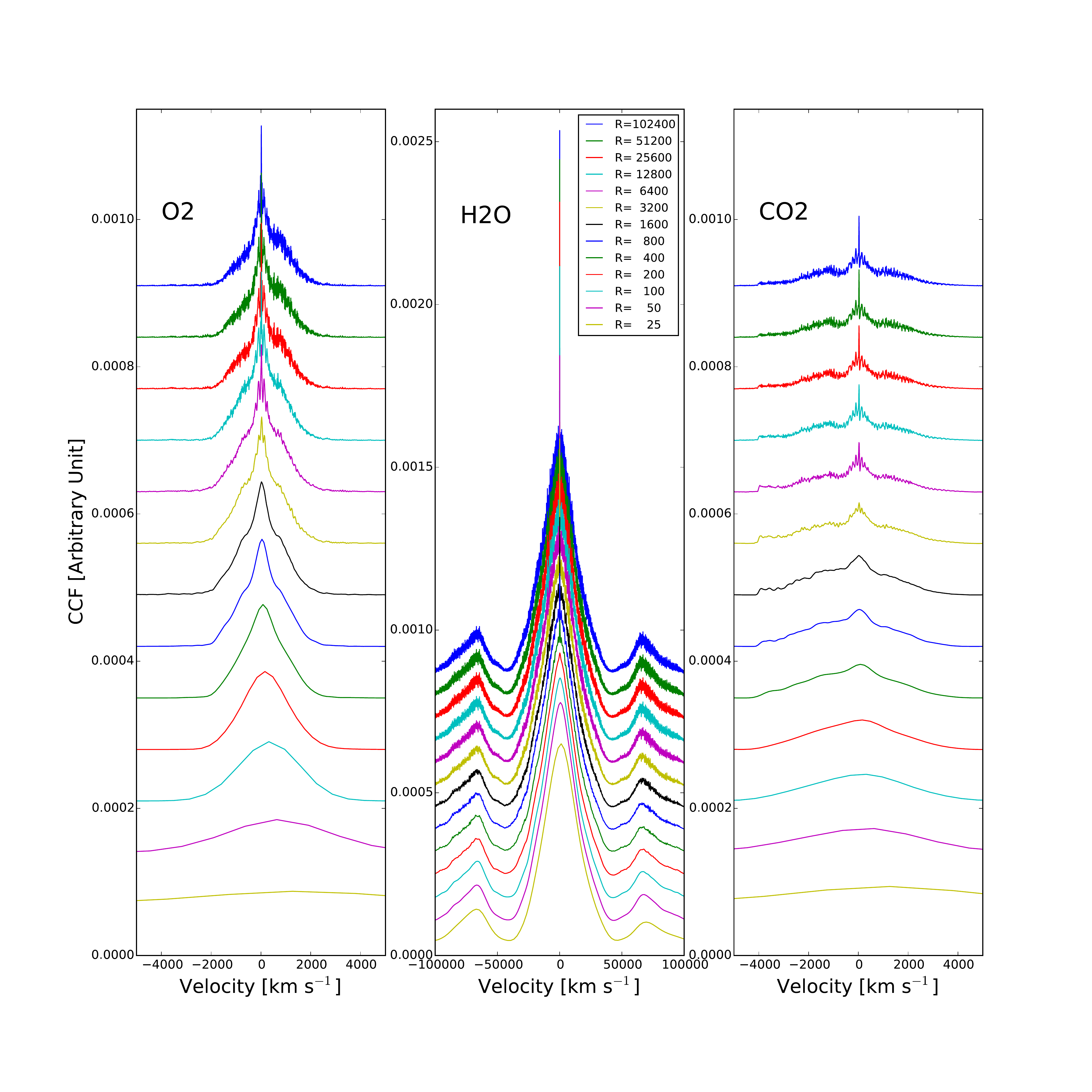}
\caption{Cross correlation functions for O$_2$, H$_2$O and CO$_2$ at different spectral resolutions.  
\label{fig:CCF}}
\end{figure*}

Fig.~\ref{fig:Space_molecular_detection} shows the CCF SNR contours in the phase space of star light suppression vs spectral resolution. An Earth-like planet with a planet/star contrast of $6.1\times10^{-11}$ can be detected at all spectral resolutions for star light suppression levels better than $2\times10^{-9}$ with a CCF SNR of 5. In the space-based photon-noise limited regime, the detectability gain of HRS is not as significant as for the ground-based case of an Earth-like planet orbiting an M star. The highest spectral resolution considered (R=102,400) increases the CCF SNR by a factor of $\sim$2 when compared to lower spectral resolutions (e.g., R=25). This can be explained by analyzing Fig.~\ref{fig:CCF}. Even at spectral resolution as low as R=25, broad H$_2$O absorption bands are resolved and this enables planet/molecular species detection with the cross correlation technique. Increasing spectral resolution helps to resolve lines and thus adds an additional fine peak on the band-resolved CCF. This additional line-resolved peak is about twice the height as the band-resolved CCF peak, which explained the factor of $\sim$2 gain in CCF SNR. 

So far, we have shown that, not surprisingly, the cross-correlation technique works on broad molecular bands at low spectral resolution.  It is interesting to note that this result comes from the fact that space observations are free of any contamination from the Earth's atmosphere. Indeed, in space, there is no need to disentangle between the signal for molecular species that the exoplanet and the Earth atmospheres might have in common. In our ground-based simulations, we apply a high-pass filter to remove the Earth's atmosphere absorption and stellar continuum low-frequency variations, which essentially removes the absorption bands from the planet signal. Therefore, ground-based simulations rely entirely on resolving absorption lines to detect the exoplanet molecular species.  

However, there are two major caveats to the photon noise limited case. First of all, the photon-noise limit cannot realistically be reached at low spectral resolution; speckle noise and its chromaticity need to be accounted for (see \S \ref{subsubsect:detnoise}).
Second, not every molecular species is readily detectable at R=25. At a star light suppression level of $10^{-10}$, O$_2$ and CO$_2$ becomes detectable at R=50 with CCF SNR of 7.1 and 3.9 respectively. This is because the absorption bands for O$_2$ and CO$_2$ are narrower than those for H$_2$O (see Fig.~\ref{fig:Molecular_absorption}). When the spectral resolutions is higher than R=50, these bands start to be resolved and the band-resolved CCF peak appears (see Fig.~\ref{fig:CCF}). 

\subsubsection{CCF at Low SNR (per pixel) Regime}

One advantage of HDC observations is the relaxation of the requirement for star light suppression level. However, in the hypothetical (unrealistic) photon-noise limited regime, the relaxation of star light suppression requirements for space-based observations is less obvious than the ground-based cases. For example, the simulation for $K_S$-band observation of M dwarf planets (Fig.~\ref{fig:Ground_K}) shows that the relaxation of star light suppression is 2-3 orders of magnitude. However, the relaxation is only a factor of $\sim$5 for the space-based case (Fig.~\ref{fig:Space_molecular_detection}) when tracing the contour of CCF SNR of 5 around $10^{-9}$ star light suppression. This is consistent with our finding in \S \ref{sec:suppression_signal} that, in the photon-noise regime, the relaxation of star light suppression level depends on the number of photons from the planet entering the instrument. 

\begin{figure}
\epsscale{1.2}
\plotone{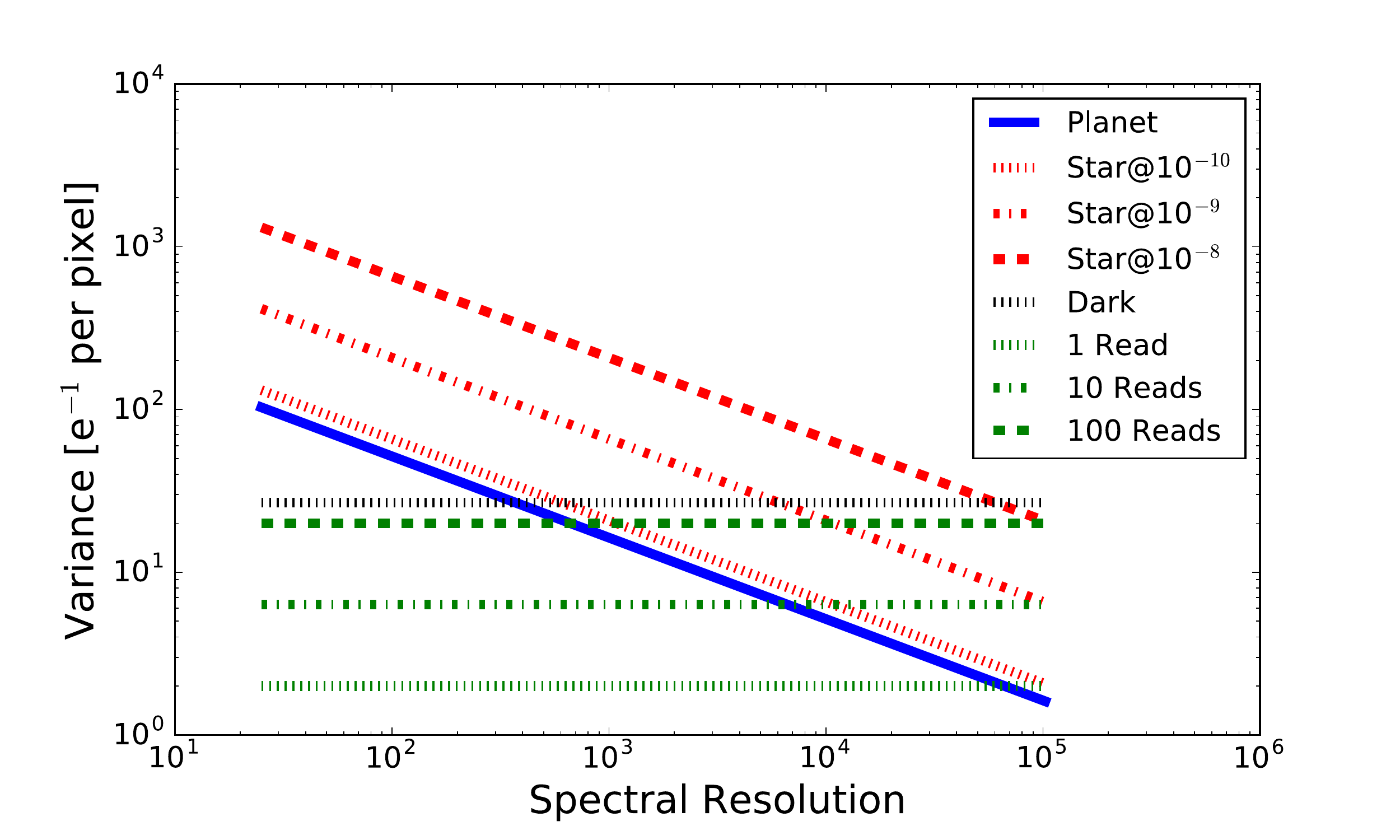}
\caption{A comparison of noise sources at different spectral resolutions for the case of 12-m space-based telescope on a Sun-Earth system at 5 pc.  
\label{fig:Space_SNR}}
\end{figure}

Space-based observations of an Earth-Sun system is extreme compared to ground-based observations of an Earth-M dwarf cases. Fig.~\ref{fig:Space_SNR} shows a comparison of noise sources at different spectral resolutions. At the highest considered spectral resolution, there are only 2-3 photons per pixel. The average SNR per pixel is $\sim$1/30 if only considering photon-noise from the star and the planet. At this low level of SNR per pixel, each absorption line is very noisy. Considering the proportionality of CCF SNR to the square root of the number of lines \citep{Snellen2015}, we only expect a modest contribution from the line-resolved CCF as shown in Fig.~\ref{fig:CCF} even if the spectral resolution is high enough to resolve individual absorption lines. 

\subsubsection{The Impact of Detector Noise, Speckle Noise and Its chromaticity}\label{subsubsect:detnoise}

\begin{figure}
\epsscale{1.2}
\plotone{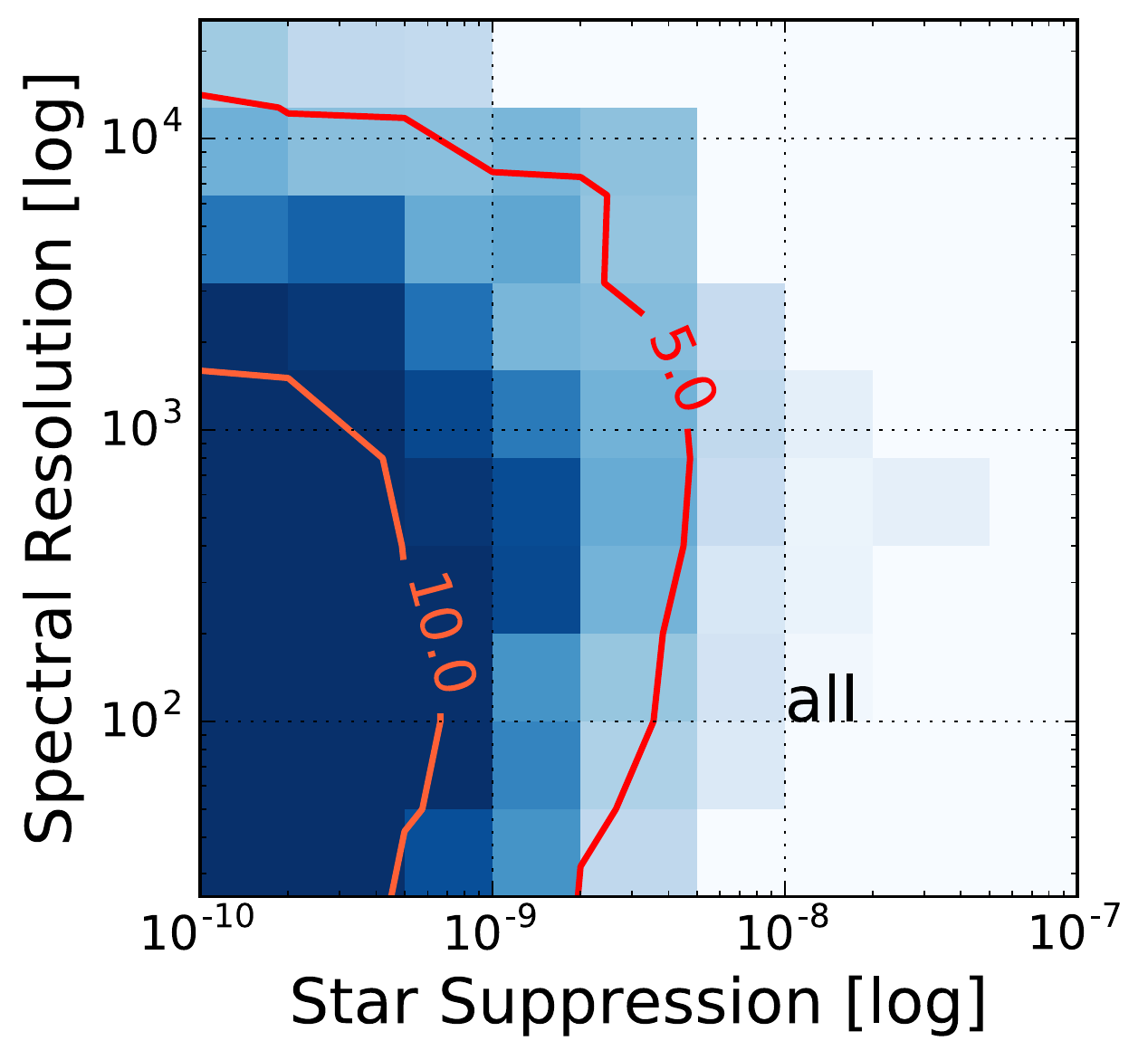}
\caption{CCF SNR contours for LUVOIR simulation in phase space of spectral resolution and star light suppression level for different molecular species. Detector noise (readout noise and dark current) is considered and values are shown in Table \ref{tab:Telescope_Instrument_Sun_Earth}. We assume 100 readouts during a 100-hr observation. 
\label{fig:LUVOIR_noise}}
\end{figure}

Fig.~\ref{fig:LUVOIR_noise} shows the impact of detector noise on planet detection (see Table~\ref{tab:Telescope_Instrument_Sun_Earth}). At low spectral resolutions, the CCF SNR contours are affected negligibly. Contours at high spectral resolutions are significantly altered due to detector noise. Fig.~\ref{fig:Space_SNR} shows that the noise contribution from dark current and readout noise for 100 readouts are comparable. When taking detector noise into account, the CCF SNR peaks at spectral resolutions lower than R=1000. This implies that future space missions should not consider extreme high resolution unless detector noise can be significantly reduced, which is an active area of research \citep{Rauscher2016}. Depending on the desired CCF SNR, the requirement for star light suppression is relaxed by 1-2 orders of magnitude compared to the astrophysical planet/star contrast, which is still very significant. The impact of detector noise on O$_2$ and H$_2$O detection is similar to the planet detection case as shown in Fig. \ref{fig:LUVOIR_noise}. However, CO$_2$ is no longer detectable after considering detector noise.  

\begin{figure}
\epsscale{1.2}
\plotone{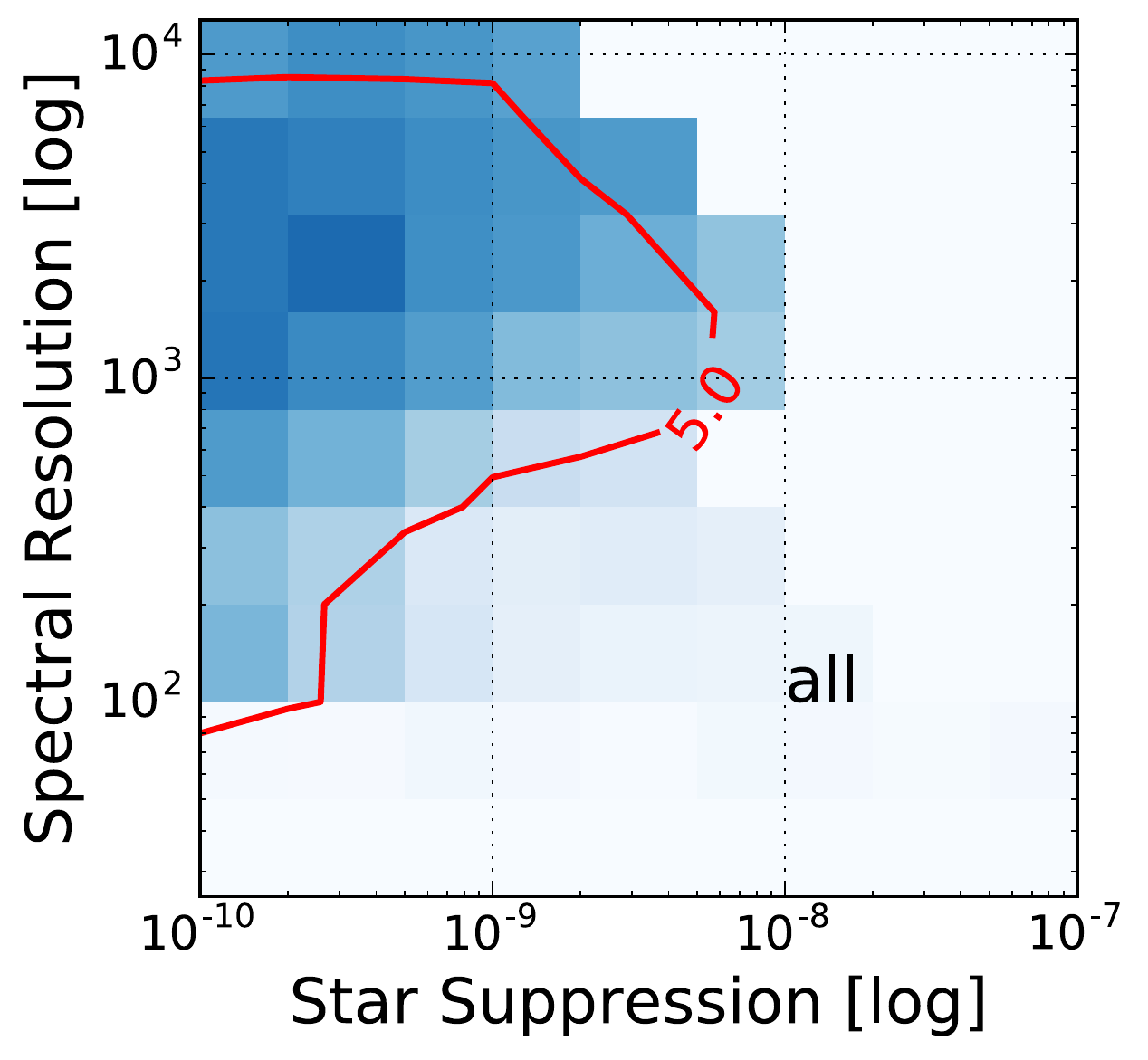}
\caption{CCF SNR contours for LUVOIR simulation in phase space of spectral resolution and star light suppression level for planet detection. Detector noise (readout noise and dark current) and speckle chromatic noise are considered. We assume 100 readouts during a 100-hr observation. 
\label{fig:LUVOIR_speckle_noise}}
\end{figure}

Fig.~\ref{fig:LUVOIR_speckle_noise} shows the CCF SNR contours including the effect of both detector noise and speckle chromatic noise. With CCF SNR greater than 5, the performance of an HDC instrument is limited by detector noise at high spectral resolution and speckle chromatic noise at low spectral resolution. We find an optimal point at R=1600 where the star light suppression requirement is relaxed to $5\times10^{-9}$, or almost 2 orders of magnitude (the Planet/Star contrast is $6.1\times 10^{-11}$).  

\subsection{Simulating HabEX Observation}

HabEx is a concept for an exoplanet direct-imaging mission with a more modest aperture than LUVOIR (4-6.5 meter). Despite a smaller aperture size, HabEx has several advantages compared to LUVOIR. First, HabEx is an exoplanet-focus mission with a much larger fraction of observing time dedicated to exoplanet search and characterization. HabEx observation can therefore afford a much longer exposure time for a single target that has a compelling case for exoplanet study. We therefore use 400 hours total exposure time in simulation, 4 times longer than what is used for LUVOIR simulation. Second, HabEx will be optimized for exoplanet direct imaging and can potentially achieve deeper star light suppression than LUVOIR. These differences between HabEx and LUVOIR need to be considered when comparing performance of HDC concepts for these two missions. 

\begin{figure*}
\epsscale{1.2}
\plotone{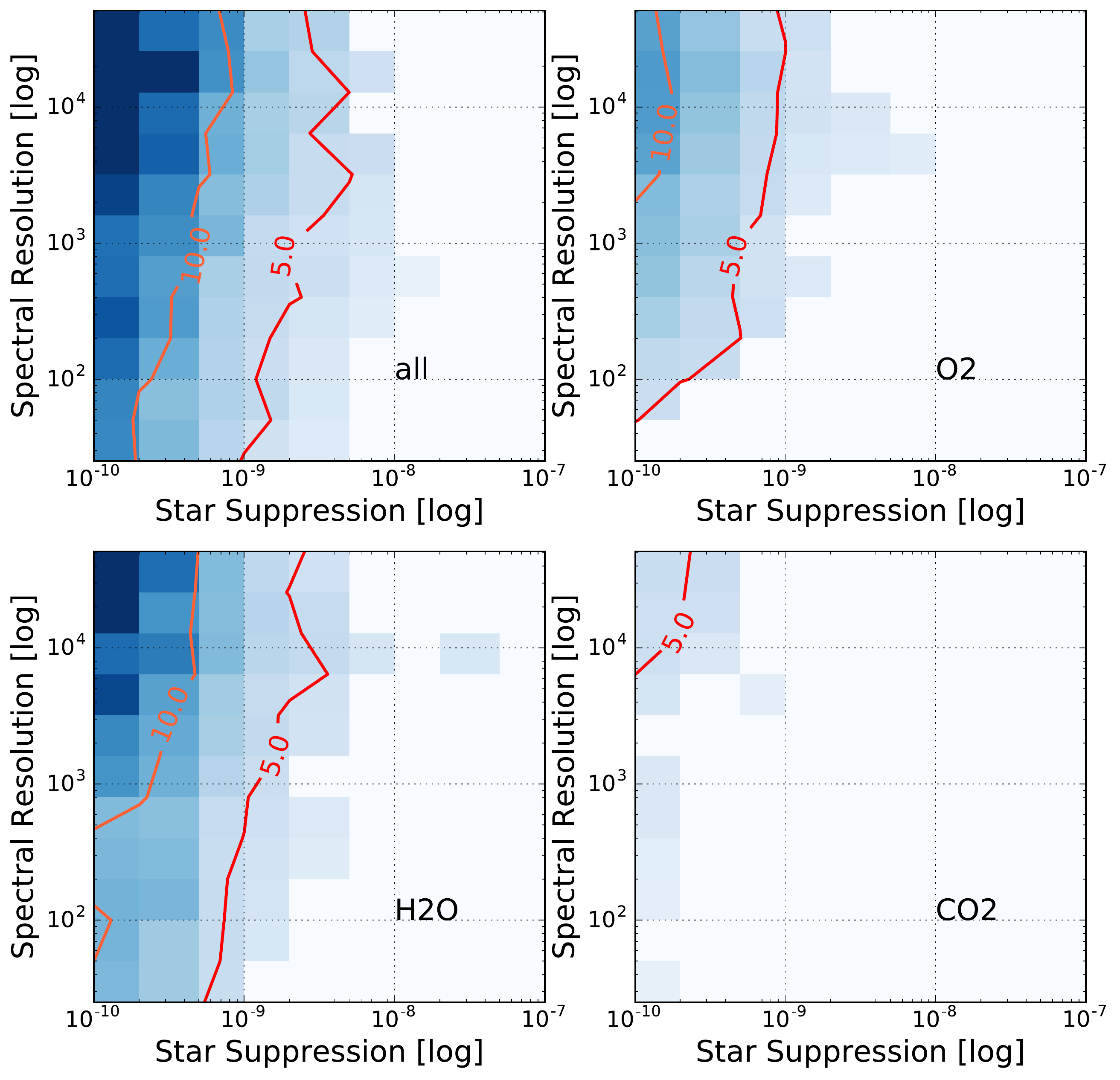}
\caption{CCF SNR contours for HabEx simulation in phase space of spectral resolution and star light suppression level for different molecular species. Detector noise (readout noise and dark current) is assumed to be zero. No speckle chromatic noise is considered. Total exposure time for HabEx simulation (i.e., 400 hr) is 4 times longer than LUVOIR simulation. This is because that LUVOIR is a general-purpose space mission and HabEx is an exoplanet-specific mission, which can afford a much longer exposure time on a single target. 
\label{fig:HabEx}}
\end{figure*}

Considering a conservative 4-meter telescope diameter, we simulate HabEx observations of a Sun-Earth system at 5 pc with total exposure time of 400 hours. Fig.~\ref{fig:HabEx} shows the CCF SNR contours vs. spectral resolutions and star light suppression levels. The results are qualitatively similar to LUVOIR simulation but with reduced CCF SNR. This is because planet signal is $\sim$2 times lower for the HabEx simulation than the LUVOIR simulation. Although we assume a 4 times longer exposure time for HabEx observations, LUVOIR has 3 times larger aperture size. 

\begin{figure}
\epsscale{1.2}
\plotone{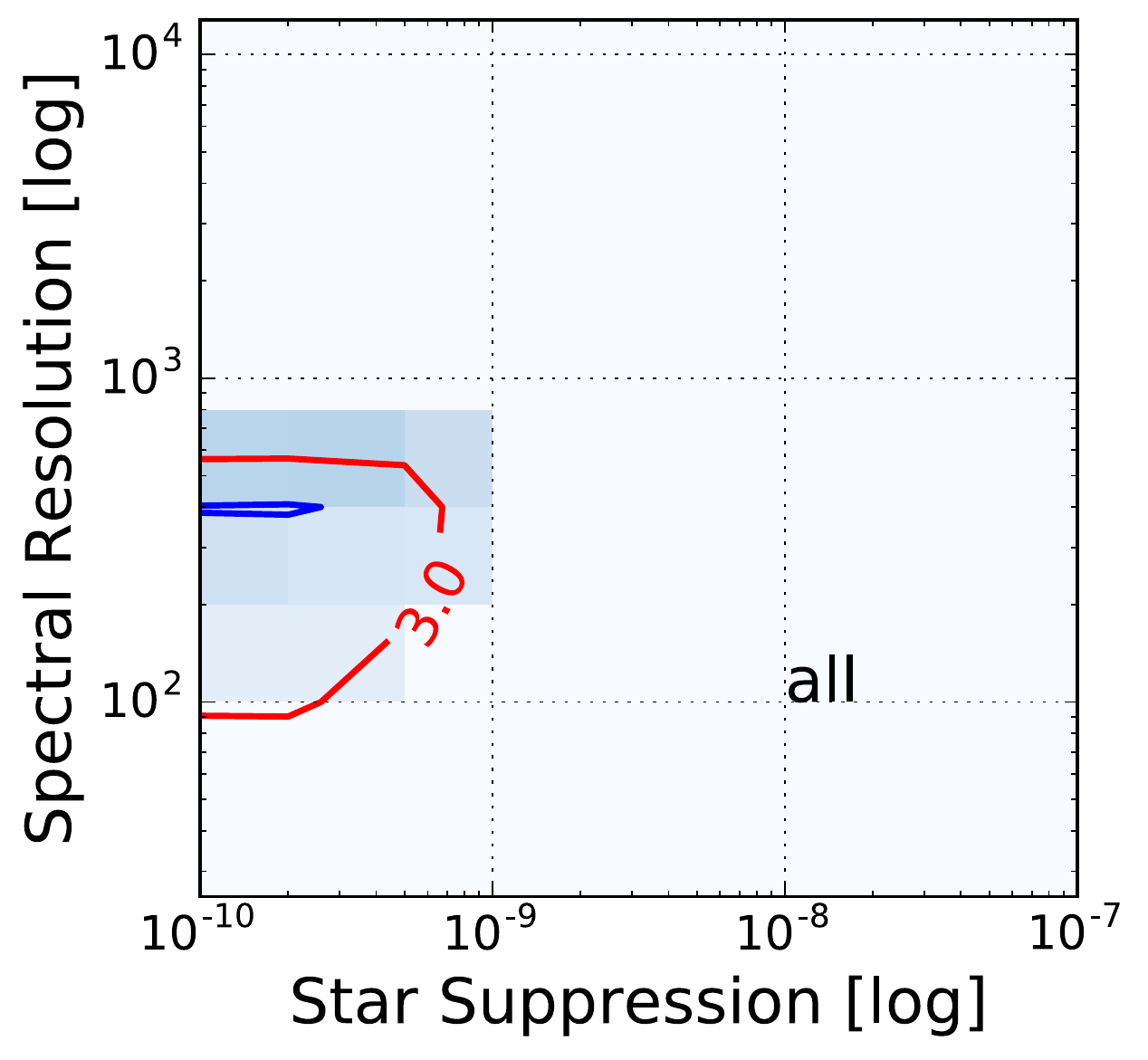}
\caption{CCF SNR contours for HabEx simulation in phase space of spectral resolution and star light suppression level for planet detection. Detector noise (readout noise and dark current) and speckle chromatic noise are considered. We assume 400 readouts during a 400-hr observation. 
\label{fig:HabEx_speckle_noise}}
\end{figure}

At C=$10^{-10}$, H$_2$O, O$_2$, and CO$_2$ start to be detected at R=25, R=50, and R=200 with CCF SNR of 9.7, 5.3, and 3.1 respectively. The highest spectral resolution we consider is R=51,200 because there is on average less than one photon per pixel for higher spectral resolutions. If considering detector noise and speckle chromatic noise, the optimal combination of spectral resolution and star light suppression for planet detection is, respectively, R=400 and C=$5\times10^{-10}$, where the CCF SNR is 4.6 (see Fig. \ref{fig:HabEx_speckle_noise}). At this combination, the relaxation of star light suppression requirement is almost a factor of $\sim$10 (the Planet/Star contrast is $6.1\times 10^{-11}$).

\section{Summary and Discussion}
\label{sec:discussion}

In this paper, we presented a framework to simulate the end-to-end performance of an HDC instrument. The pipeline intakes spectra of planets and stars and considers atmospheric transmission and background emission if applicable. With a realistic assumption of coronagraphic and spectroscopic system performance, the pipeline simulates observed and reduced planet spectra with reasonable noise sources including photon-noise from planet,  star, the Earth's atmosphere background emission, detector noise and speckle chromatic noise. The pipeline also simulates the subsequent spectral analysis, such as detecting a planet and the molecular species in its atmosphere using the cross correlation method. The pipeline can be used for the trade study of future ground-based and space-based missions dedicated to the search and characterization of exoplanets. We provide a few representative test cases: (1) observations of currently known directly imaged planets (i.e., HR 8799 e and 51 Eri b) with the 10-m Keck telescope; (2) observations of Proxima Cen b and an Earth-like planet around a M dwarf at 5 pc with a 30-m class ground-based telescope; (3) observations of an Earth-Sun system at 5 pc with 4-m and 12-m space-based telescopes. These simulations are valuable in terms of understanding the power and limitation of the HDC technique.  

\subsection{Lessons Learned From Simulations For Currently-Known Directly-Imaged Exoplanets}

We applied the pipeline to currently known directly imaged planets HR 8799 e and 51 Eri b. We studied the feasibility of detecting such planets and characterizing the composition of their atmospheres using KPIC, a Keck HDC instrument under development. We summarize our findings as follows.

\begin{itemize}
  \item The CCF SNR is not always photon-noise limited. Other factors that limit CCF SNR include the intrinsic structure of the CCF and the mismatch between the observed spectrum and the template that is used in the cross correlation (Fig. \ref{fig:HR8799e} and Fig. \ref{fig:51Erib}).
  \item The mismatch spectrum case yields the lowest CCF SNR. This result highlights the importance of planet spectrum modeling in the cross correlation method. However, the mismatched spectrum case also represents an opportunity for atmosphere retrieval by varying model parameters to maximize the CCF peak. 
  \item Multi-band observation is necessary in order to fully characterize the chemical composition of a planet. We considered three molecular species (CO, H$_2$O, and CH$_4$) and demonstrated the optimal band for detecting them can be different (Fig.~\ref{fig:HR8799_mol_JHKL} and \ref{fig:51Erib_mol_JHKL}).
  \item The increased sensitivity makes the HDC technique suitable for planet searches. For example, ~\citet{Lovis2016} considered an integral field unit (IFU) formed by a bundle of 7 hexagonal single-mode fibers to search for the exact location of Proxima Cen b. Similarly,~\citet{Rains2016} proposed a 3 by 3 fiber-based IFU for planet search and characterization. METIS~\citep{Brandl2014}, one of the three first light instruments for E-ELT, will provide $L$ and $M$ band IFU capability for HDC observations. The multiplexing capability increases the effective field of view and relaxes the requirement for pointing and tracking stability at the expense of detector size or wavelength coverage. 
\end{itemize}

\subsection{Lessons Learned From Simulations For Systems with an Earth-Like Planet}

HDC simulations for the observation of an Earth-like planet was pioneered by~\citet{Sparks2002}. This important topic was later on explored by a few groups~\citep{Riaud2007, Kawahara2014, Snellen2015, Lovis2016}. The present study builds on previous work in terms of simulation methodology. We thoroughly explored the parameter space of spectral resolution and star light suppression level for both space-based and ground-based observations. 


\subsubsection{Ground-based Observations}

Here we summarize our findings from the simulations of the ground-based observations of Earth-like planets in the habitable zone of M dwarfs. 

\begin{itemize}
  \item High spectral resolution allows the star light suppression requirements for detection and characterization to be relaxed by 2-3 orders of magnitude. Using the HDC technique, we find that the minimum star light suppression level is 10$^{-4}$ for the 10-$\sigma$ detection of Proxima Cen b with ground-based 30-m class telescopes.
  \item In addition to Proxima Cen b, extremely large ground-based telescopes can be used to study M dwarf planet systems further away (e.g., 5 pc). Given the abundance of M dwarfs~\citep{Cantrell2013} and planets around M dwarfs~\citep{Dressing2015} in the solar neighborhood within 5 pc, the prospect of studying planets around cool stars is promising.
  \item The performance of an HDC instrument depends on the planet signal and its relative strength with respect to other noise sources (Fig. \ref{fig:Ground_SNR}). In order to reach the full potential of an HDC instrument and push for higher sensitivity at lower levels of star light suppression, increasing signal throughput from the planet is the key.
  \item The dominating noise source for ground-based HDC observation with 30-m class telescope is the photon noise from leaked stellar light (for 100-hr observation, see Fig. \ref{fig:Ground_SNR}). Detector and sky background noise are $\sim$5 times lower than the dominating noise source at $10^{-6}$ star light suppression level.
  \item High spectral resolution (R$\sim$100,000) and deep star light suppression ($\sim10^{-8}$) offers a unique opportunity to study exo-terrestrial atmosphere at unprecedentedly high SNR (Fig.~\ref{fig:Ground_J}-\ref{fig:Ground_K}) although severe technological hurdles need to be overcome. 

\end{itemize}

\subsubsection{Comparing to Previous Results}

\citet{Kawahara2014} predicted that $10^{-4}$ and $10^{-5}$ star light suppression levels are required for the 3$\sigma$ detection of H$_2$O. We ran a simulation with a similar setup as theirs and find that H$_2$O can be detected with a CCF SNR of 6.8 at star light suppression level of $5\times10^{-4}$. We suspect the the factor of 2 difference in detection significance may be attributed to different approach in calculating the CCF SNR. 

~\citet{Snellen2015} investigated the detectability of a short-period super Earth around Proxima Cen and concluded that the planet can be detected with a significance of 10 for wavelength coverage from 0.6 to 0.9 $\mu$m with an HDC instrument on the E-ELT. Following the details in their paper, we found a CCF SNR of 4.0 for such a super Earth. We note that the star light suppression level assumed in~\citet{Snellen2015} is $\sim3.3\times10^{-4}$ whereas we found that a star light suppression level of  $1\times10^{-4}$ is the minimal requirement for detection. The difference in CCF SNR and starlight suppression requirement can be explained by the spectra used in cross correlation. ~\citet{Snellen2015} considered both reflected stellar lines and planetary molecular absorption lines whereas we considered only planetary molecular absorption lines. Within the 0.6 to 0.9 $\mu$m wavelength coverage, the reflected stellar lines contribute more to CCF peak than the planetary molecular absorption lines. Not considering the reflected stellar lines results in a lower CCF SNR than ~\citet{Snellen2015}. However, in $J$, $H$ and $K$ band, the planetary molecular absorption lines contribute much more to CCF peak than the reflected stellar lines. Therefore, considering the reflected stellar lines does not significantly improve CCF SNR in near infrared wavelengths. 

~\citet{Lovis2016} investigated the potential of SPHERE+EXPRESSO on VLT to search and characterize Prox Cen b. They found that the planet can be detected at 5-$\sigma$ with a total of 240h integration time. We adopted their values in our pipeline and found that the CCF SNR is 6.8-9.2 for a star light suppression level between 1/5000-1/2000.  

\subsubsection{Space-based Observation}
Here we summarize our findings from the simulations of the space-based observations of  Earth-like planets in the habitable zone of a solar-type star.

\begin{itemize}
  \item For a 12-m space-based telescope, an Earth-like planet can be detected at all spectral resolutions (R=25-102,400) for star light suppression levels better than $2\times10^{-9}$ with a CCF SNR of 5 (Fig. \ref{fig:Space_molecular_detection}). For a 4-m space-based telescope, the CCF SNR reduces because of the smaller aperture size (Fig. \ref{fig:HabEx}). 
  \item The number of photons from the planet entering the instrument is critical. As the aperture increases from the 4 meters (HabEx-like telescope) to 12 meters (LUVOIR-like telescope), we find a significant increase of CCF SNR for all molecular species for a fixed exposure time of 100 hours. \item While the 12-m LUVOIR concept has a 3 times larger aperture size than the 4-m HabEx concept, HabEx can afford a much longer exposure time and can potentially achieve deeper star light suppression because it is focused and optimized for exoplanet study. These differences between HabEx and LUVOIR need to be considered when comparing performance of HDC concepts.
  \item Space-based observations can operate at low spectral resolution without the concern of contamination by the Earth's atmosphere (speckle chromaticity might be worse though, see next point). Planet or molecular species can be detected by their absorption bands at spectral resolutions as low as R=25. In contrast, for ground-based observation, we apply a high-pass filter to remove the Earth's atmosphere absorption and stellar continuum low-frequency variations, which essentially removes the absorption bands from the planet signal. Therefore, ground-based observations rely entirely on resolving absorption lines for detection. That is the regime where HRS comes into play as a critical component. 
  \item The performance of an HDC instrument is limited by detector noise at high spectral resolution and speckle chromatic noise at low spectral resolution (Fig. \ref{fig:LUVOIR_speckle_noise} and \ref{fig:HabEx_speckle_noise}).
  \item Future space missions should not consider extreme high resolution unless detector noise can be significantly reduced.

\end{itemize}



\subsection{Future works}

In a future paper, we wish to establish a quantitative relationship between planet signal and relaxation of the requirements for star light suppression. In particular, we want to answer quantitatively how the gain by HRS in an HDC instrument changes with planet signal in the presence of various noise sources. In addition, we will make the simulations more realistic by considering details in echelle spectroscopic data reduction. One outstanding question is how to preserve extremely weak planet signal (a few to hundreds photons per pixel) at every step of the data reduction and spectral analysis. 

\noindent{\it Acknowledgements}

\bibliography{mybib_JW_DF_PH5}

\clearpage

\newpage
\input{Telescope_Instrument.tex} 
\newpage
\input{HR8799e.tex} 
\newpage
\input{51Erib.tex} 
\newpage
\input{Telescope_Instrument_Mdwarf_Earth.tex} 
\newpage
\input{ProxCenb.tex} 
\newpage
\input{Mdwarf_Earth.tex} 
\newpage
\input{Telescope_Instrument_Sun_Earth.tex} 
\newpage
\input{Sun_Earth.tex} 
\newpage
\end{document}

%% file: Telescope_Instrument.tex

\begin{deluxetable}{ccc}
\tablewidth{0pt}
\tablecaption{Telescope and instrument parameters for simulated observations of HR 8799 e and 51 Eri b .\label{tab:telescope_instrument}}
\tablehead{
\colhead{\textbf{Parameter}} &
\colhead{\textbf{Value}} &
\colhead{\textbf{Unit}} \\
}

\startdata

Telescope aperture & 10.0 & m \\
Spectral resolution & 37500 & \nodata \\
$J$ band spectral range & 1.143 - 1.375 & $\mu$m \\
$H$ band spectral range & 1.413 - 1.808 & $\mu$m \\
$K$ band spectral range & 1.996 - 2.382 & $\mu$m \\
$L^\prime$ band spectral range & 3.420 - 4.120 & $\mu$m \\
Exposure time & 3600 & second \\
Fiber angular diameter & 1.0 & $\lambda$/D \\
Wavefront correction residual$^\ast$ & 260 & nm \\
Telescope+instrument throughput$^{\ast\ast}$ & 10\% & \nodata \\
Readout noise & 3.0 & e$^{-}$ \\
Dark current & 0.01 & e$^{-}$ s$^{-1}$

\enddata

\tablecomments{$\ast$: Private communication with Peter Wizinowich. $\ast\ast$: This throughput is for $K$ band. Throughputs for other bands are scaled with Strehl ratio. }

\end{deluxetable}

%% file: HR8799e.tex

\begin{deluxetable}{cccc}
\tablewidth{0pt}
\tablecaption{HR 8799 and planet e.\label{tab:HR8799e}}
\tablehead{
\colhead{\textbf{Parameter}} &
\colhead{\textbf{Value}} &
\colhead{\textbf{Unit}} &
\colhead{\textbf{References}} \\
}

\startdata

\multicolumn{4}{l}{\textbf{Star}} \\
Effective temperature (T$_{\rm{eff}}$) & 7193 & K & ~\citet{Baines2012} \\
Surface gravity ($\log g$) & 4.03 & cgs & ~\citet{Baines2012} \\
Distance & 39.40 & pc & ~\citet{vanLeeuwen2007} \\
V$\sin i$ & 37.5 & km s$^{-1}$ & ~\citet{Kaye1998} \\
Inclination ($i$)$^{\ast}$ & $>\sim$ 40 & degree & ~\citet{Wright2011a} \\
Radial velocity & -11.5 & km s$^{-1}$  & ~\citet{Gontcharov2006} \\
\multicolumn{4}{l}{\textbf{Planet}} \\
Effective temperature (T$_{\rm{eff}}$) & 1100-1650 & K & ~\citet{Bonnefoy2015} \\
Surface gravity ($\log g$) & 3.5-4.1 & cgs & ~\citet{Bonnefoy2015} \\
Metallicity ([M/H]) & 0.0-0.5 & dex & ~\citet{Bonnefoy2015} \\
V$\sin i$$^{\ast\ast}$ & $<$40.0 & km s$^{-1}$ & ~\citet{Konopacky2013} \\
Inclination ($i$) & 28 & degree & ~\citet{Soummer2011} \\
Semi-major axis ($a$) & 14.94-20.44 & AU & ~\citet{Zurlo2015} \\
Radial velocity$^{\ast\ast\ast}$ & -11.5 & km s$^{-1}$  & ~\citet{Gontcharov2006} \\
Angular separation & 0.38-0.52 & arcsec & ~\citet{Zurlo2015} \\
Angular separation in $J$ & 14.6-20.2 & $\lambda$/D & ~\citet{Zurlo2015} \\
Angular separation in $H$ & 11.4-15.7 & $\lambda$/D & ~\citet{Zurlo2015} \\
Angular separation in $K_S$ & 8.4-11.5 & $\lambda$/D & ~\citet{Zurlo2015} \\
Angular separation in $L^\prime$ & 4.7-6.9 & $\lambda$/D & ~\citet{Zurlo2015} \\
Planet/star contrast in $J$ & $2.0\times10^{-6}$ & \nodata & \nodata \\
Planet/star contrast in $H$ & $1.0\times10^{-5}$ & \nodata & \nodata \\
Planet/star contrast in $K_S$ & $3.8\times10^{-5}$ & \nodata & \nodata \\
Planet/star contrast in $L^\prime$ & $2.1\times10^{-4}$ & \nodata & \nodata \\
\enddata

\tablecomments{$^{\ast}$: We adopt 40 degree in simulations. $^{\ast\ast}$: We assume a rotational velocity of 15 km s$^{-1}$. $^{\ast\ast\ast}$: Assumed to be the same as HR 8799. }

\end{deluxetable}

%% file: 51Erib.tex

\begin{deluxetable}{cccc}
\tablewidth{0pt}
\tablecaption{51 Eri and planet b.\label{tab:51Erib}}
\tablehead{
\colhead{\textbf{Parameter}} &
\colhead{\textbf{Value}} &
\colhead{\textbf{Unit}} &
\colhead{\textbf{References}} \\
}

\startdata

\multicolumn{4}{l}{\textbf{Star}} \\
Effective temperature (T$_{\rm{eff}}$)$^{\ast}$ & 7400 & K & \nodata \\
Surface gravity ($\log g$)$^{\ast\ast}$ & 4.0 & cgs & \nodata \\
Distance & 29.40 & pc & ~\citet{Macintosh2014} \\
Rotational velocity & 50.0 & km s$^{-1}$ & \nodata \\
Inclination ($i$) & 40.0 & degree & \nodata \\
Radial velocity & -12.6 & km s$^{-1}$  & ~\citet{Gontcharov2006} \\
\multicolumn{4}{l}{\textbf{Planet}} \\
Effective temperature (T$_{\rm{eff}}$) & 550-750 & K & ~\citet{Macintosh2014} \\
Surface gravity ($\log g$) & 3.5 & cgs & ~\citet{Macintosh2014} \\
Rotational velocity & 15.0 & km s$^{-1}$ & \nodata \\
Inclination ($i$) & 45 & degree & \nodata \\
Projected separation ($a$) & 13.2 & AU & ~\citet{Macintosh2014} \\
Radial velocity$^{\ast\ast\ast}$ & -12.6 & km s$^{-1}$  & ~\citet{Gontcharov2006} \\
Angular separation & 0.45 & arcsec & ~\citet{Macintosh2014} \\
Angular separation in $J$ & 17.3 & $\lambda$/D & ~\citet{Macintosh2014} \\
Angular separation in $H$ & 13.5 & $\lambda$/D & ~\citet{Macintosh2014} \\
Angular separation in $K_S$ & 9.9 & $\lambda$/D & ~\citet{Macintosh2014} \\
Angular separation in $L^\prime$ & 5.8 & $\lambda$/D & ~\citet{Macintosh2014} \\
Planet/star contrast in $J$ & $2.6\times10^{-6}$ & \nodata & \nodata \\
Planet/star contrast in $H$ & $1.1\times10^{-6}$ & \nodata & \nodata \\
Planet/star contrast in $K_S$ & $1.7\times10^{-6}$ & \nodata & \nodata \\
Planet/star contrast in $L^\prime$ & $2.7\times10^{-5}$ & \nodata & \nodata \\
\enddata

\tablecomments{$^{\ast}$ and $^{\ast\ast}$: Based on F0IV spectral estimation from ~\citet{Macintosh2014}. $^{\ast\ast\ast}$: Assumed to be the same as 51 Eri. }

\end{deluxetable}

%% file: Telescope_Instrument_Mdwarf_Earth.tex

\begin{deluxetable}{ccc}
\tablewidth{0pt}
\tablecaption{Telescope and instrument parameters for M dwarf planets (Proxima Cen b and a M dwarf planet system at 5 pc).\label{tab:Telescope_Instrument_Mdwarf_Earth}}
\tablehead{
\colhead{\textbf{Parameter}} &
\colhead{\textbf{Value}} &
\colhead{\textbf{Unit}} \\
}

\startdata

Telescope aperture & 30.0 & m \\
Telescope+instrument throughput & 10\% & \nodata \\
Wavefront correction error floor & 200 & nm \\
Spectral resolution & varied & \nodata \\
$J$ band spectral range & 1.143 - 1.375 & $\mu$m \\
$H$ band spectral range & 1.413 - 1.808 & $\mu$m \\
$K$ band spectral range & 1.996 - 2.382 & $\mu$m \\
Exposure time & 100 & hour \\
Fiber angular diameter & 1.0 & $\lambda$/D \\
Readout noise & 0.0 or 2.0 & e$^{-}$$^\ast$ \\
Dark current & 0.0 or 0.002 & e$^{-}$ s$^{-1}$$^\ast$

\enddata

\tablecomments{$\ast$: Based on H2RG detector specification~\citep{Blank2012} }

\end{deluxetable}

%% file: ProxCenb.tex

\begin{deluxetable}{ccc}
\tablewidth{0pt}
\tablecaption{Proxima Centauri b planet system.\label{tab:ProxCenb}}
\tablehead{
\colhead{\textbf{Parameter}} &
\colhead{\textbf{Value}} &
\colhead{\textbf{Unit}} \\
}

\startdata

\multicolumn{3}{l}{\textbf{Star}} \\
Effective temperature$^{\ast}$ (T$_{\rm{eff}}$) & 3050 & K  \\
Mass & 0.12 & $M_\odot$  \\
Radius & 0.14 & $R_\odot$  \\
Surface gravity ($\log g$) & 5.0 & cgs  \\
Metallicity ([M/H]) & 0.0 & dex  \\
Distance & 1.295 & pc  \\
V$\sin i$ & $<$1 & km s$^{-1}$  \\
Inclination ($i$) & 20 & degree  \\
Radial velocity & -22.4 & km s$^{-1}$   \\
\multicolumn{3}{l}{\textbf{Planet}} \\
Effective temperature (T$_{\rm{eff}}$) & 234 & K  \\
V$\sin i$$^{\ast\ast}$ &0.014 & km s$^{-1}$  \\
Inclination ($i$) & 20 & degree  \\
Semi-major axis ($a$) & 0.05 & AU  \\
Radius & 1.0 & R$_\oplus$ \\
Radial velocity & 22.2 & km s$^{-1}$   \\
Illuminated Area & 0.5 &  \nodata \\
Planet/Star Contrast & $1.6\times10^{-7}$ & \nodata \\
Angular separation & 38.6 & mas  \\
Angular separation in $J$ & 4.5 & $\lambda$/D  \\
Angular separation in $H$ & 3.5 & $\lambda$/D  \\
Angular separation in $K_S$ & 2.6 & $\lambda$/D  \\
\enddata

\tablecomments{$\ast$: All values are from ~\citet{Escude2016}. We use 3000 K in simulation. $\ast\ast$: We assume that the planet is tidally locked. }

\end{deluxetable}

%% file: Mdwarf_Earth.tex

\begin{deluxetable}{ccc}
\tablewidth{0pt}
\tablecaption{An M dwarf and an Earth-like Planet.\label{tab:Mdwarf_Earth}}
\tablehead{
\colhead{\textbf{Parameter}} &
\colhead{\textbf{Value}} &
\colhead{\textbf{Unit}} \\
}

\startdata

\multicolumn{3}{l}{\textbf{Star}} \\
Effective temperature (T$_{\rm{eff}}$) & 3500 & K  \\
Mass & 0.5 & $M_\odot$  \\
Radius & 0.5 & $R_\odot$  \\
Surface gravity ($\log g$) & 4.5 & cgs  \\
Metallicity ([M/H]) & 0.0 & dex  \\
Distance & 5.0 & pc  \\
V$\sin i$ & 2.7 & km s$^{-1}$  \\
Inclination ($i$) & 20 & degree  \\
Radial velocity & 15.0 & km s$^{-1}$   \\
\multicolumn{3}{l}{\textbf{Planet}} \\
Effective temperature (T$_{\rm{eff}}$) & 300 & K  \\
Surface gravity ($\log g$) & 3.0 & cgs  \\
V$\sin i$ &0.017 & km s$^{-1}$  \\
Inclination ($i$) & 20 & degree  \\
Semi-major axis ($a$) & 0.1 & AU  \\
Radius & 1.0 & R$_\oplus$ \\
Radial velocity & 20.0 & km s$^{-1}$   \\
Illuminated Area & 0.5 &  \nodata \\
Planet/Star Contrast & $6.2\times10^{-9}$ & \nodata \\
Angular separation & 20.0 & mas  \\
Angular separation in $J$ & 2.3 & $\lambda$/D  \\
Angular separation in $H$ & 1.8 & $\lambda$/D  \\
Angular separation in $K_S$ & 1.3 & $\lambda$/D  \\
\enddata


\end{deluxetable}

%% file: Telescope_Instrument_Sun_Earth.tex

\begin{deluxetable}{ccc}
\tablewidth{0pt}
\tablecaption{Telescope and instrument parameters for LUVOIR or HabEx.\label{tab:Telescope_Instrument_Sun_Earth}}
\tablehead{
\colhead{\textbf{Parameter}} &
\colhead{\textbf{Value}} &
\colhead{\textbf{Unit}} \\
}

\startdata

Telescope aperture & 4.0 or 12.0 & m \\
Telescope+instrument throughput & 10\% & \nodata \\
Wavefront correction error floor & 5 & nm \\
Spectral resolution & varied & \nodata \\
Spectral range & 0.5 - 1.7 & $\mu$m \\
Exposure time & 400 or 100 & hour \\
Fiber angular diameter & 1.0 & $\lambda$/D \\
Readout noise & 0.0 or 2.0$^{\ast}$ & e$^{-}$$^{\ast}$ \\
Dark current & 0.0 or 0.002 or $5.5\times10^{-6}$$^{\ast\ast}$ & e$^{-}s^{-1}$ 

\enddata

\tablecomments{$\ast$: Based on H2RG detector specification~\citep{Blank2012} and e2v CCD specification. $\ast\ast$: Used for O$_2$ detection.}

\end{deluxetable}

%% file: Sun_Earth.tex

\begin{deluxetable}{ccc}
\tablewidth{0pt}
\tablecaption{A Sun-Earth System at 5 pc.\label{tab:Sun_Earth}}
\tablehead{
\colhead{\textbf{Parameter}} &
\colhead{\textbf{Value}} &
\colhead{\textbf{Unit}} \\
}

\startdata

\multicolumn{3}{l}{\textbf{Star}} \\
Effective temperature (T$_{\rm{eff}}$) & 5800 & K \\
Mass & 1.0 & $M_\odot$  \\
Radius & 1.0 & $R_\odot$  \\
Surface gravity ($\log g$) & 4.5 & cgs  \\
Metallicity ([M/H]) & 0.0 & dex  \\
Distance & 5.0 & pc  \\
Rotational velocity & 2.0 & km s$^{-1}$  \\
Inclination ($i$) & 50 & degree  \\
Radial velocity & 0,0 & km s$^{-1}$   \\
\multicolumn{3}{l}{\textbf{Planet}} \\
V$\sin i$$^{\ast\ast\ast}$ &0.5 & km s$^{-1}$  \\
Inclination ($i$) & 50 & degree  \\
Semi-major axis ($a$) & 1.0 & AU  \\
Radius & 1.0 & R$_\oplus$ \\
Radial velocity & 20.4 & km s$^{-1}$   \\
Illuminated Area & 0.5 &  \nodata \\
Planet/Star Contrast & $6.1\times10^{-11}$ & \nodata \\
Angular separation & 200.0 & mas  \\
Angular separation at 1 $\mu$m for 12-m aperture& 11.6 & $\lambda$/D  \\
Angular separation at 1 $\mu$m for 4-m aperture & 3.9 & $\lambda$/D  \\
\enddata


\end{deluxetable}